\documentclass{article}
\usepackage{arxiv}
\usepackage[utf8]{inputenc} 
\usepackage[T1]{fontenc}    
\usepackage{url}            
\usepackage{booktabs}       
\usepackage{amsfonts}       
\usepackage{nicefrac}       
\usepackage{microtype}      
\usepackage{lipsum}
\usepackage[backref=page]{hyperref}
\usepackage{amsmath,amsfonts,amsthm,amssymb,bm,bbm} 
\usepackage{MnSymbol}

\makeatother
\usepackage{dirtytalk}
\usepackage{float}
\usepackage{siunitx}
\usepackage{booktabs}
\usepackage{multirow}
\usepackage{tabulary}
\usepackage{graphicx}
\usepackage{tabularx}
\usepackage{rotating}
\usepackage{adjustbox}

\usepackage[normalem]{ulem} 

\newcommand{\beginsupplement}{%
        \setcounter{table}{0}
        \renewcommand{\thetable}{S\arabic{table}}%
        \setcounter{figure}{0}
        \renewcommand{\thefigure}{S\arabic{figure}}%
     }

\title{Drug prescription clusters in the UK Biobank: An assessment of drug-drug interactions and patient outcomes in a large patient cohort}

\author{
  Kyriakos Schwarz
  \thanks{Biomedical Informatics, University Hospital of Zurich, Zurich, Switzerland}
  \thanks{ETH AI Center, Zurich, Switzerland}
  \\
  Department of Quantitative Biomedicine\\
  University of Zurich\\
  Schmelzbergstrasse 26, Zurich, CH \\
  \texttt{kyriakos.schwarz@uzh.ch} \\
   \And
  Daniel Trejo Banos\footnotemark[1] \\
  Department of Quantitative Biomedicine\\
  University of Zurich\\
  Schmelzbergstrasse 26, Zurich, CH \\
  \texttt{daniel.trejobanos@uzh.ch} \\
   \And
  Giulia Rathmes\footnotemark[1] \\
  Department of Quantitative Biomedicine\\
  University of Zurich\\
  Schmelzbergstrasse 26, Zurich, CH \\
  \texttt{giulia.rathmes@uzh.ch} \\
   \And
  Michael Krauthammer
  \footnotemark[1]
  \footnotemark[2]
  \\
  Department of Quantitative Biomedicine\\
  University of Zurich\\
  Schmelzbergstrasse 26, Zurich, CH \\
  \texttt{michael.krauthammer@uzh.ch} \\
}

\begin{document}
\maketitle

\begin{abstract}

 In recent decades, there has been an increase in polypharmacy, the concurrent administration of multiple drugs per patient. Studies have shown that polypharmacy is linked to adverse patient outcomes and there is interest in elucidating the exact causes behind this observation. In this paper, we are studying the relationship between drug prescriptions, drug-drug interactions (DDIs) and patient mortality. Our focus is not so much on the number of prescribed drugs, the typical metric in polypharmacy research, but rather on the specific combinations of drugs leading to a DDI. To learn the space of real-world drug combinations, we first assessed the drug prescription landscape of the UK Biobank, a large patient data registry. We observed distinct drug constellation patterns driven by the UK Biobank participants' disease status. We show that these drug prescription clusters matter in terms of the number and types of expected DDIs, and may possibly explain observed differences in health outcomes.

\end{abstract}

\vspace*{0.5cm}

\clearpage

\section*{Introduction}

\begin{figure}[ht]
  \centering
      \includegraphics[width=0.5\textwidth]{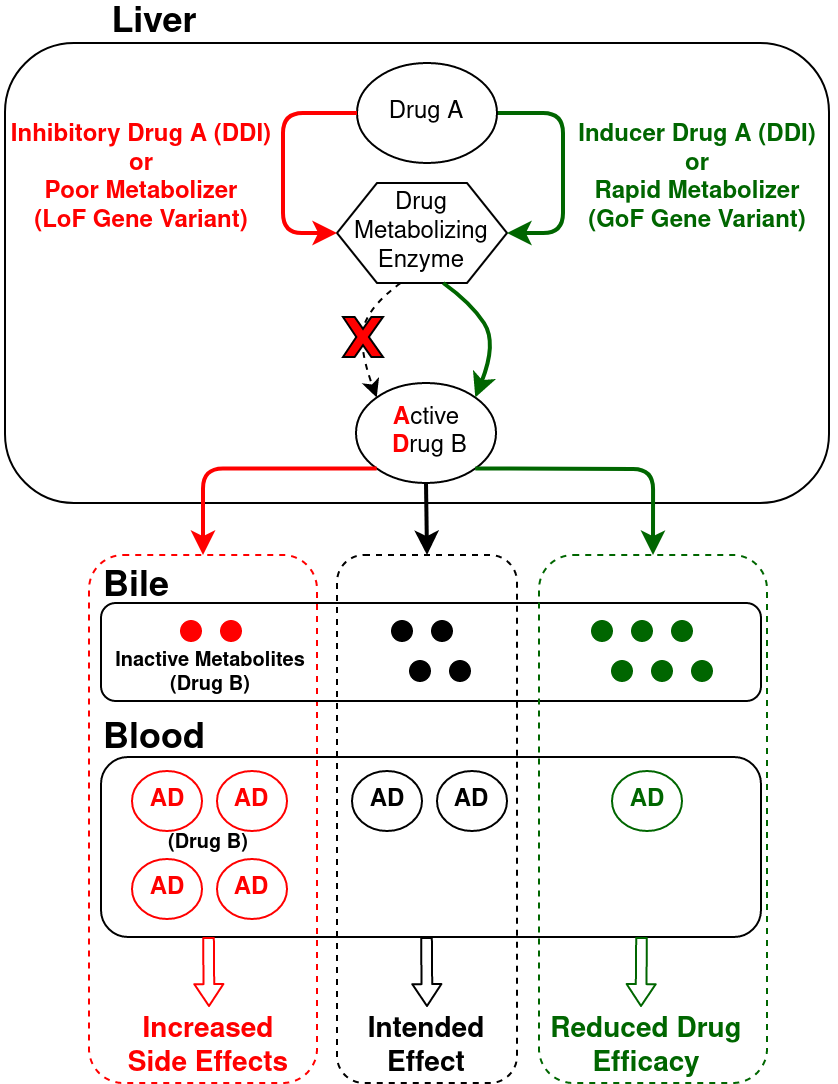}
      \caption{Connection of the effects of DDIs and GVs.\label{fig:DrugBloodSE}}
\end{figure}

Over the last decades, patients have been prescribed an increasing number of drugs leading to a rise in polypharmacy, the concurrent use of five or more drugs \cite{kantor_trends_2015}. Polypharmacy has been associated with a high risk of adverse outcomes in patients, including mortality, hospitalizations, and falls \cite{hilmer_effects_2009}. Much research has been devoted to improve our understanding of these polypharmacy-related adverse events. One prominent line of research is to study the role of drug-drug interactions (DDIs). For example, the concurrent use of drugs may lead to the activation or inactivation of another drug's metabolizing enzyme. This effect may be enhanced by existing genetic variation in drug-metabolizing genes, i.e., genotypes that result in slower or faster drug metabolism and may lead to increased side effects or reduced efficacy of a particular drug (Figure \ref{fig:DrugBloodSE}).

Inherently, polypharmacy exposes patients to several possible DDIs. However, current polypharmacy research often ignores the specific combination of drugs prescribed to patients. Instead, it focuses on summary measures such as total drugs used and their effect on particular health outcomes \cite{maher2014clinical}. Therefore, it is paramount to study polypharmacy on the level of individual drugs, which necessitates a careful examination of the combination of drugs involved and the effect of these combinations on potential DDIs and other health processes.

Large biomedical cohorts, such as the UK Biobank \cite{sudlow_uk_2015}, offer vast opportunities for health outcome research. For example, Anastopoulos et al. \cite{anastopoulos_multi-drug_2021} utilized UK Biobank patient characteristics (age, sex, weight, etc.), as well as multi-drug features for the prediction of hospitalizations and mortality. Zemedikun et al. \cite{zemedikun_patterns_2018} analyzed the patterns of multimorbidity and disease clusters within the UK Biobank and highlighted pairs of diseases that have an increased likelihood of co-occurrence. Furthermore,  McQueenie et al. \cite{mcqueenie_multimorbidity_2020} assessed the influence of multimorbidity and polypharmacy on the risk of COVID-19. 

To our knowledge, the specific drug prescription patterns in polypharmacy have not been extensively studied in the UK Biobank. We are particularly interested in examining whether specific drug combinations may lead to additional negative (or even positive) health effects in the UK Biobank participants. Our first objective was, therefore, to better understand the real-world topology of concurrent drug prescriptions in UK Biobank and, as a second objective, to study the participants' outcomes with respect to their drug prescriptions. To this end, we performed a study across $328'185$ UK Biobank participants in order to: 1) Identify clusters of patients with similar drug prescriptions (i.e., discover "drug constellations") and 2) study these drug constellations with regard to patient outcomes. To achieve this aim, we mapped UK Biobank drug information into "\textbf{Meta-Drugs}", categories that provide a semantically harmonized view of the participants' drug prescriptions. We also integrated additional UK Biobank data, including demographic, clinical and genetic information, in order to better separate the effect of drug prescriptions from other participant features, i.e., minimize confounder effects.

\clearpage

\section*{Methods}
\subsection*{Datasets}
\subsubsection*{1. UK Biobank main dataset}
The UK Biobank is a large-scale biomedical database and research resource, containing genetic and health information from roughly $500'000$ UK participants \cite{sudlow_uk_2015}. For the purposes of this study, we focused on the participants' characteristics as listed in Table \ref{table:ukb_data_fields}.

\begin{table}[ht]
\begin{center}
    \normalsize
\begin{tabular}{r|l|l}
Data Field & Description & Usage                                                                      \\\hline
31 & Sex & As is \\
21003 & Age when attended assessment centre & As is \\
21000 & Ethnic Background & Mapped to broader ethnic categories \\
189 & Townsend deprivation index at recruitment & As is \\
20001 & Cancer code & Merged with Data Field 20002 \\
20002 & Non-cancer illness code & Merged with Data Field 20001 \\
20003 & Treatment/medication code & Mapped to drug ingredients and ATC codes\\
2296 & Falls in the last year & As is \\
46 & Left hand strength & Max value of  Data Fields 46,47 - normalized by sex (31) \\
47 & Right hand strength & Max value of Data Fields 46,47 - normalized by sex (31) \\
924 & Usual walking pace & As is \\
20116 & Smoking status & As is \\
40007 & Age at death & Calculated five-year mortality  \\
40001 & Primary cause of death & As is \\
\end{tabular}
    \end{center}
    \caption{UK Biobank Data Fields.\label{table:ukb_data_fields}}
\end{table}

For the data fields that were not used "As is" (see Table \ref{table:ukb_data_fields}) we provide a data field processing description below.

\subsubsection*{2. Medication information}

We accessed information from a previous UK Biobank study \cite{wu_genome-wide_2019} for mapping participants' medications to Anatomical Therapeutic Chemical Classification System (ATC) codes and active ingredients. This allowed us to map $6'745$ unique medication mentions in the UK Biobank to $691$ unique active ingredients or combinations thereof, as multiple medication codes may map to the same active ingredient. For example, the original UK Biobank descriptions for Aspirin (e.g., "aspirin 75mg tablet", "disprin direct dispersible tablet", "micropirin 75mg e/c tablet", etc.) are all mapped to the drug ingredient "Acetylsalicylic Acid". Those active ingredients were further mapped to $275$ unique ATC codes, as multiple ingredients may share the same ATC code. For instance, the ingredients "Ditazol", "Cloricromen", "Picotamid" and "Acetylsalicylic Acid" all share the ATC code "B01AC".

\subsubsection*{3. Phenotypes of drug metabolizing enzymes}

We obtained a dataset from a previous UK Biobank study \cite{mcinnes_pharmacogenetics_2021} (Return dataset $3388$) which contains presumptive phenotypes (e.g., fast or slow metabolizer) of $14$ drug-metabolizing enzymes for roughly $50'000$ UK Biobank participants. The genes included are: \textit{CFTR, CYP2B6, CYP2C19, CYP2C9, CYP2D6, CYP3A5, CYP4F2, DPYD, IFNL3, NUDT15, SLCO1B1, TPMT, UGT1A1} and \textit{VKORC1}.

\subsubsection*{4. Drug-Drug Interactions}

Through "Therapeutics Data Commons" \cite{Huang2021tdc}, we obtained the "DrugBank Multi-Typed DDI" dataset, which contains $191'808$ DDI pairs for $1'706$ drugs and $60$ distinct DDI types. 

We utilized $11$ out of the $60$ DDI types for additional, individual DDI type analysis (Table \ref{table:ddi_types}). These  selected DDI types are linked to well described physiological effects, which is essential for interpreting the associations between DDIs and patient outcomes.

\begin{table}[ht]
\begin{center}
    \normalsize
\begin{tabular}{r|l}
\textbf{DDI type ID} & \textbf{DDI type description}
\\\hline
29 & Drug A may decrease the diuretic activities of Drug B \\
6 & Drug A may increase the anticoagulant activities of Drug B \\
37 & Drug A may decrease the antihypertensive activities of Drug B \\
32 & Drug A may increase the sedative activities of Drug B \\
83 & Drug A may increase the hypokalemic activities of Drug B \\
9 & Drug A may increase the hypoglycemic activities of Drug B \\
20 & Drug A may increase the QTc-prolonging activities of Drug B \\
10 & Drug A may increase the antihypertensive activities of Drug B \\
57 & Drug A may increase the nephrotoxic activities of Drug B \\
60 & Drug A may increase the hypotensive activities of Drug B \\
27 & Drug A may increase the neuroexcitatory activities of Drug B \\
\end{tabular}
    \end{center}
    \caption{Selected DDI types for additional, individual DDI type analysis.\label{table:ddi_types}}
\end{table}

\subsection*{Processing of Data Fields}

Ethnic background values (Data Field $21000$) were mapped to broader categories according to Table \ref{table:ethnic_back_mapping}. Cancer codes and Non-cancer illness codes (Data Fields $20001$ and $20002$) were merged into a single field. Treatment/medication codes (Data Field $20003$) were mapped to active ingredients and ATC codes according to the description above (see section "Medication Information"). "Left" and "Right hand strength" (Data Fields $46$ and $47$) were merged by the maximum of the two values and subsequently normalized by sex (Data Field $31$). We calculated five-year mortality by subtracting the "Age when attended assessment centre" (Data Field $21003$) from Age at death (Data Field $40007$). An additional field, "Drug counts", was constructed by counting the active ingredients per participant. 

\subsection*{Learning of Meta-Drugs}

As we are interested in common prescription patterns that are informed by drug indication - and not just drug ingredients, we decided to group the unique drug ingredients ($n=691$) in the UK Biobank according to drug indication. We refer to those groups as \textbf{Meta-Drugs}. For example, calcium channel blockers and ACE inhibitors were grouped into a common Meta-Drug group "hypertension".

For this purpose, we first constructed a unified patient disease list. We merged the "Non-cancer illness code" (diseases excluding cancer) and "Cancer code" (all cancer-related diseases) fields, obtaining a list of $525$ unique medical conditions (diseases) in total. We then counted the number of occurrences of participants with specific drug-disease combinations, thus constructing a Drug-Disease count matrix. This matrix was manually filtered by a medical expert in order to only map drugs to diseases for which they have a medical indication. The final resulting matrix contained $589$ drug ingredients and $167$ drug-targeted diseases (Figure \ref{fig:clustering_overview} - A). 

For the $589$ drugs, pairwise correlations were calculated, which were subsequently grouped together by hierarchical clustering. Hence, $90$ Meta-Drugs were identified (Figure \ref{fig:clustering_overview} - A). Figure \ref{fig:meta-drug-counts} gives an overview of the number of drugs per Meta-Drug.

\subsection*{Clustering of Participants}

\begin{figure}[ht]
  \centering
      \includegraphics[width=1.0\textwidth]{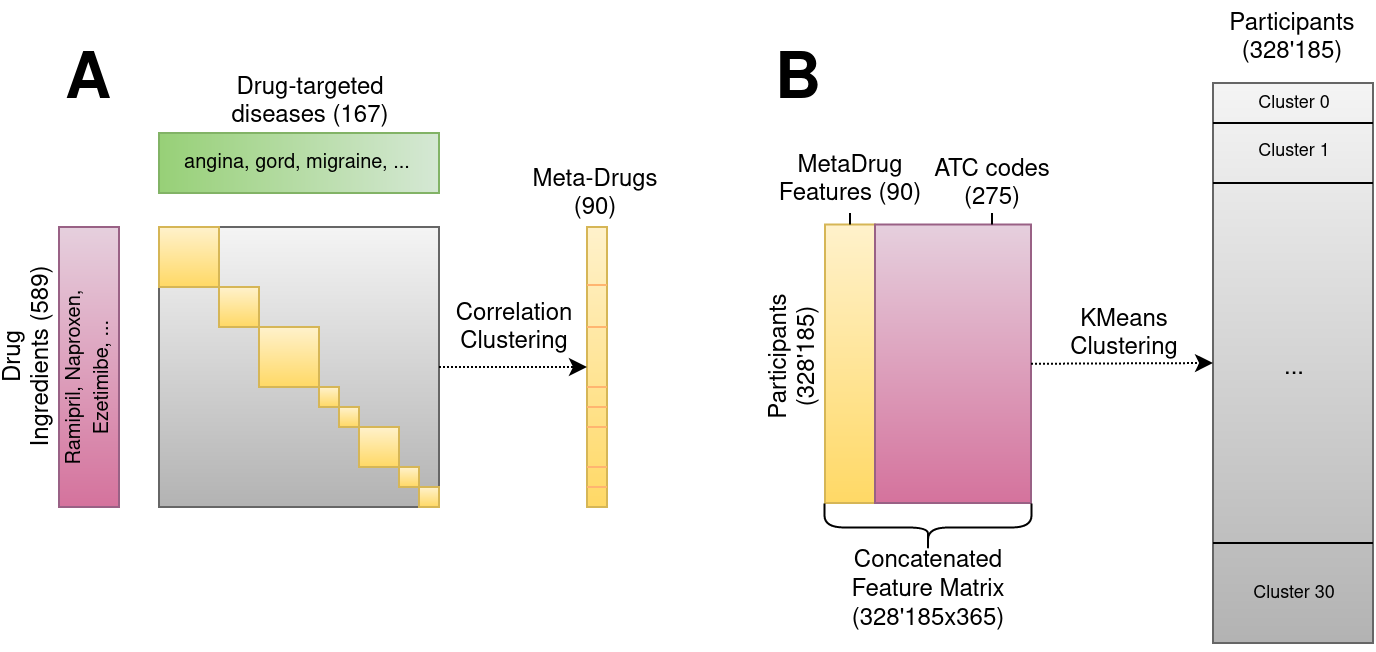}
      \caption{Overview of Meta-Drug learning and Participant clustering. \textbf{(A)} A matrix of co-occurring drugs and diseases of the $328'185$ participants is constructed. The drug-disease matches are filtered by a medical expert. A correlation clustering is performed on the filtered drug-disease matrix in order to identify $90$ "\textbf{Meta-Drugs}", i.e., groups of drugs that treat similar diseases. \textbf{(B)} A concatenated participant-Meta-Drug and participant-ATC feature matrix is constructed and subsequently utilized as input to KMeans clustering. $31$ main clusters of participants are identified according to their individual drug prescriptions.\label{fig:clustering_overview}}
\end{figure}

We identified $328'185$ participants that were administered at least one of the $589$ active drug ingredients. For each participant, we constructed a feature vector consisting of the participant's Meta-Drugs ($1$-hot encoded binary vector of length $90$) concatenated with the participant's ATC codes ($1$-hot encoded binary vector of length $275$). The total feature vector length is $365$ (Figure \ref{fig:clustering_overview} - B).

\begin{figure}[ht]
  \centering
      \includegraphics[width=\textwidth,height=\textheight,keepaspectratio]{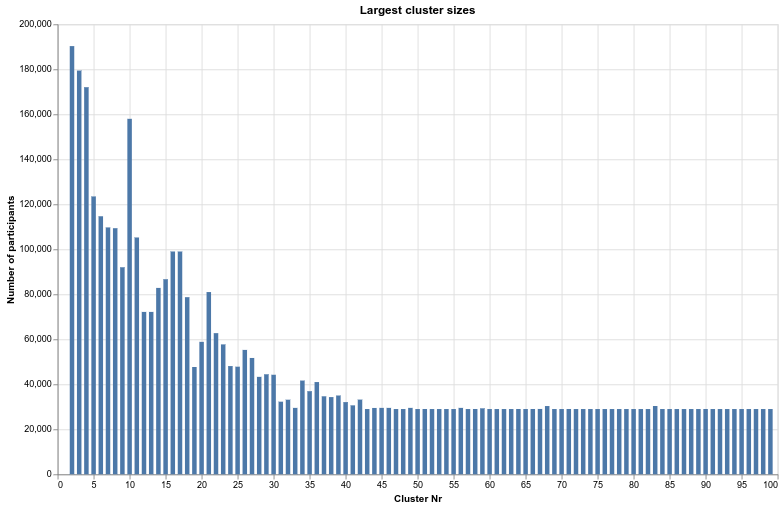}
      \caption{Distribution of the largest cluster for each K for the main clusters.\label{fig:c31-largest-clusters}}
\end{figure}

\begin{figure}[ht]
  \centering
      \includegraphics[width=\textwidth,height=\textheight,keepaspectratio]{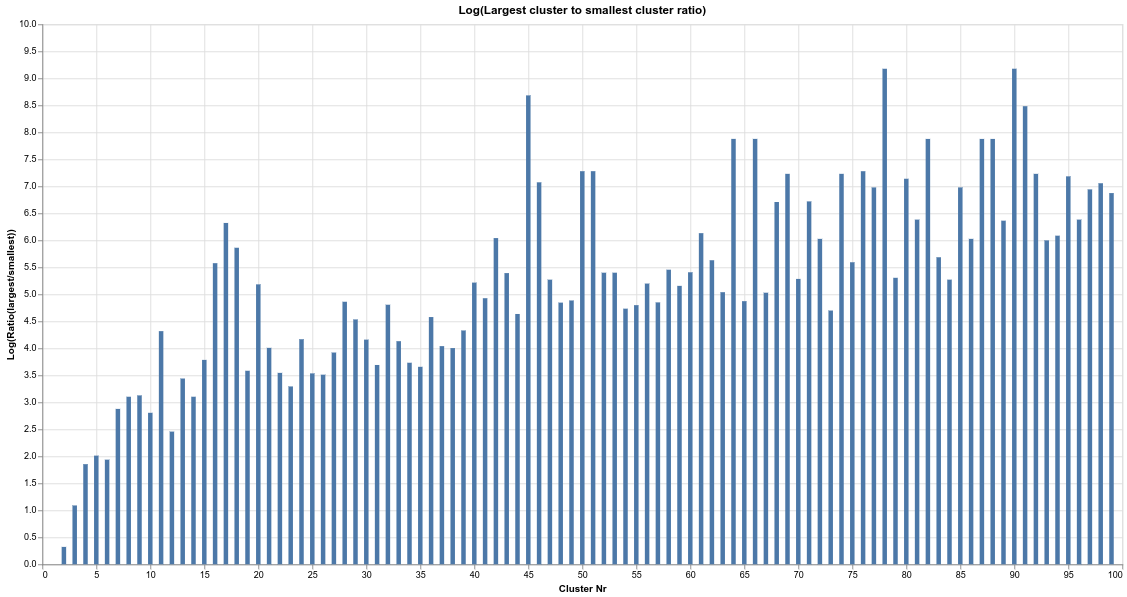}
      \caption{Distribution of the log largest to smallest ratio for each K for the main clusters.\label{fig:c31-log-ratio}}
\end{figure}

Subsequently, we performed a K-Means clustering on the normalized $328'185 \times 365$ feature matrix, with $K \in [2,99]$. In order to determine the optimal value $K$ we calculated the K-Means inertia scores ("elbow method"), which suggested a value of $20 \leq K \leq 40$ (Figure \ref{fig:c31-elbow-method}). Furthermore, in order to avoid having large discrepancies in cluster sizes we calculated the largest cluster size, as well as the largest-to-smallest cluster ratio for each $K$ (Figures \ref{fig:c31-largest-clusters} and \ref{fig:c31-log-ratio}). According to these criteria, we selected $K = 31$, which resulted in more homogeneous cluster sizes compared to other values of $K$.

Henceforth, we refer to the main clusters as $C0$ to $C30$.

\subsection*{Statistical comparison of cluster attributes}

In order to compare cluster attributes we performed statistical tests as per Table \ref{table:stattests}. Depending on the attribute type, we performed a two-proportions Z-test for proportional values (equation \ref{eq:z-statistic}), a two-sample Kolmogorov-Smirnov test for continuous values (equation \ref{eq:KS-statistic}), or calculated the Wasserstein distance for discrete values (equation \ref{eq:Wasserstein}). While the Chi-square test is typically utilized for comparison of discrete distributions, our data did not fit the criteria for this test. Instead, the Wasserstein distance was calculated. The utilized statistical tests are described in detail in the supplementary methods.

\begin{table}[ht]
\begin{center}
    \normalsize
\begin{tabular}{l|l}
\textbf{Cluster attribute} & \textbf{Comparative measure}
\\\hline
Sex &  Z-test \\
Smoking status &  Z-test  \\
Usual walking pace &  Z-test  \\
DDI type counts &  Z-test  \\
Falls in the last year &  Z-test  \\
Five-year mortality &  Z-test  \\
Age &  KS-test  \\
Townsend deprivation index at recruitment &  KS-test  \\
Normalized hand strength &  KS-test  \\
Drug counts & Wasserstein distance \\
\end{tabular}
    \end{center}
    \caption{Statistical assessment and comparison of cluster attributes.\label{table:stattests}}
\end{table}

\clearpage

\section*{Results}

In this study, we pursued two main objectives: to learn about drug prescription constellations in the UK Biobank and to study these constellations with regard to the participants' outcomes. To pursue these objectives and to get a comparable view of participants' prescriptions, we first attempted to harmonize drug representations in the UK Biobank. 

\subsection*{Harmonization of drug representations in the UK Biobank: learning of Meta-Drugs}

\begin{figure}[ht]
  \centering
      \includegraphics[width=\textwidth,height=\textheight,keepaspectratio]{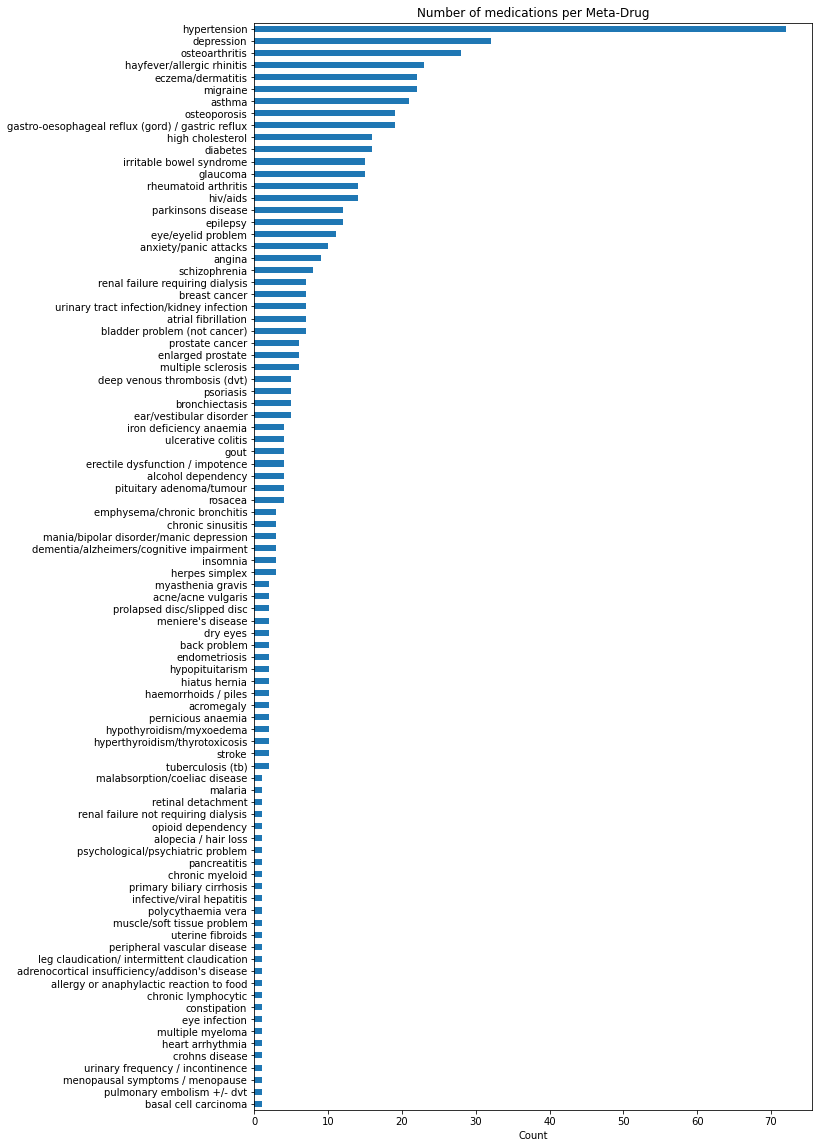}
      \caption{Number of drugs per Meta-Drug.\label{fig:meta-drug-counts}}
\end{figure}

The medications in the UK Biobank are listed as simple text without any reference to a standardized drug database or identifier. Thus, we improved this drug representation by mapping these text entries to ($1$) harmonized ATC codes and ($2$) "\textbf{Meta-Drugs}", i.e., a high-level grouping of drugs by indication (both procedures are described in detail in the Methods section, see Figure \ref{fig:clustering_overview}). This grouping resulted in $90$  Meta-Drugs (starting with 589 unique drug entries in the UK Biobank), with some Meta-Drugs subsuming up to $72$ different drug ingredients (see Figure \ref{fig:meta-drug-counts}).

As an example, the original UK Biobank descriptions for Aspirin (e.g., "aspirin 75mg tablet", "disprin direct dispersible tablet", "micropirin 75mg e/c tablet", etc.) were all mapped to the corresponding ATC code "B01AC". Furthermore, this drug ingredient was mapped to the Meta-Drug "meta-angina" (i.e., group of drugs that treat coronary heart disease) which includes the following other drugs: Verapamil, Diltiazem, Isosorbide Dinitrate, Isosorbide Mononitrate, Nicorandil, Nitroglycerin and Clopidogrel.

\begin{figure}[ht]
  \centering
      \includegraphics[width=\textwidth]{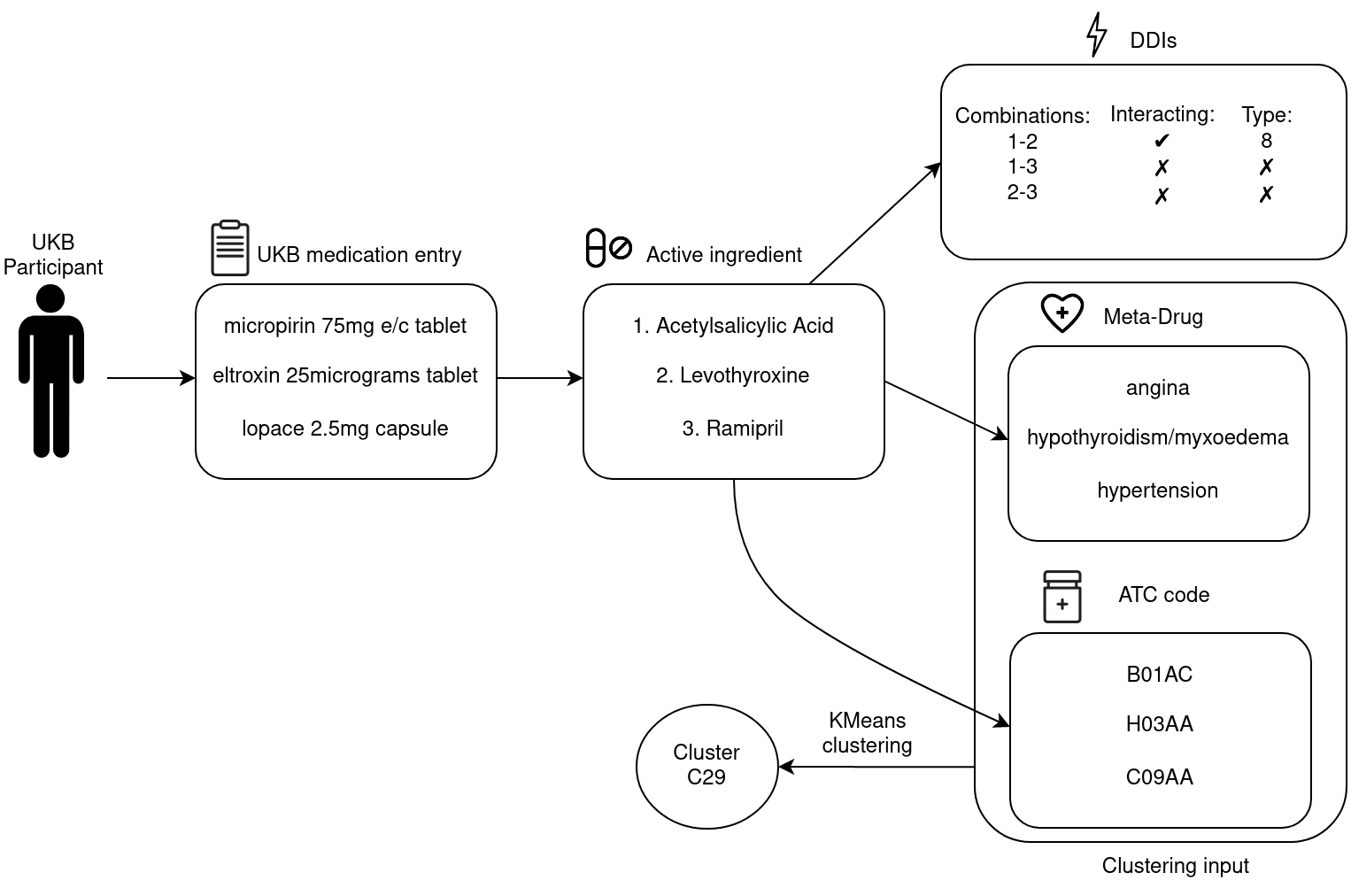}
      \caption{UK Biobank participant example - from medication entries to clusters of drug prescriptions. The UK Biobank medication entries are first mapped to the active drug ingredients, which are then mapped to  "\textbf{Meta-Drugs}" (subsuming drugs treating similar diseases) and ATC codes (drug classification system). These mappings are then used as an input for clustering UK Biobank participants into drug constellations using KMeans clustering. The figure further shows our procedure for detecting the presence of possible DDIs among a participant's list of prescribed drugs. For all active ingredients, the number of DDIs and their types are calculated according to DDI information from DrugBank (see Methods section). For each drug ingredient combination (e.g., drug combination 1-2) the participant is assigned a DDI type (e.g., type $8$ - "The therapeutic efficacy of Drug A can be increased when used in combination with Drug B").
      \label{fig:participant-example}}
\end{figure}

\subsection*{Drug prescription constellations in the UK Biobank}

\begin{figure}[ht]
  \centering
      \includegraphics[width=\textwidth,height=\textheight,keepaspectratio]{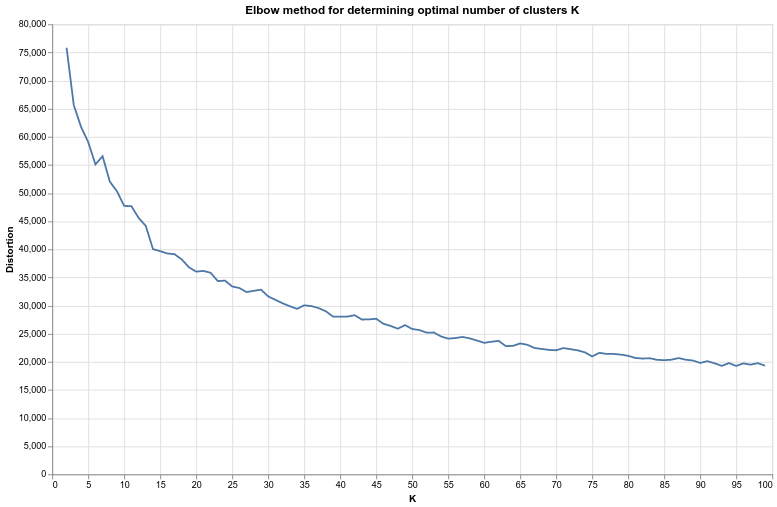}
      \caption{Elbow method for determining the optimal number of clusters, K, for the main clusters.\label{fig:c31-elbow-method}}
\end{figure}

After K-Means clustering (see Methods section for details) of the $328'185$ participants based on their list of medications (ATC codes and Meta-Drugs), we identified $K = 31$ main clusters that reveal the diverse drug prescription constellations present in this cohort. The optimal number of clusters ($31$) was determined by inertia scores (Figure \ref{fig:c31-elbow-method}). 
The characteristics of the resulting clusters are listed in Tables \ref{table:c31_summary_stats} (main cluster characteristics), \ref{table:c31_top5_meds} (top five medications per cluster), \ref{table:c31_top5_medconds} (top five medical conditions per cluster), \ref{table:c31_top5_metadrugs} (top five Meta-Drugs per cluster), \ref{table:c31_top3_drugcombs} (top three medication combinations per cluster) and \ref{table:c31_top3_metacombs} (top three Meta-Drug combinations per cluster).

\subsubsection*{Cluster characteristics}

As shown in Table \ref{table:c31_summary_stats}, cluster sizes range from $786$ ($C11$) to $31'124$ ($C22$) participants. The average age, number of drugs and five-year mortality ranges from $52.9$ to $61.4$ years, from $1.0$ to $6.3$ drugs and from $0.6 \%$ to $5.2 \%$, respectively. 

The clusters reflect a shared clinical situation of the participants. For example, $C20$ features hyper- and hypothyroidism - typically seen in younger female participants \cite{vanderpump1995incidence} - as the top two diseases. Indeed, the cluster features more women ($85.8\%$) and younger individuals (average age $56.1$ years) than most other clusters. $C26$ has the highest mortality, is comprised mainly of women ($98.5\%$) and features breast cancer as the top diagnosis and Tamoxifen, a selective estrogen receptor modulator, as the top medication. $C27$ consists mainly of older men ($73.6\%$ men, average age $58.6$ years) with enlarged prostate (top diagnosis) and Tamsulosin, Finasteride and Dutasteride - all medications for benign prostatic hyperplasia - among the top five medications. $C29$ features gastro-oesophageal reflux and a PPI (proton pump inhibitor for acid reduction) as the top disease and medication, respectively. 

\subsubsection*{Meta-Drug distributions across the clusters}

\begin{figure}[ht]
  \centering
      \includegraphics[width=\textwidth,height=\textheight,keepaspectratio]{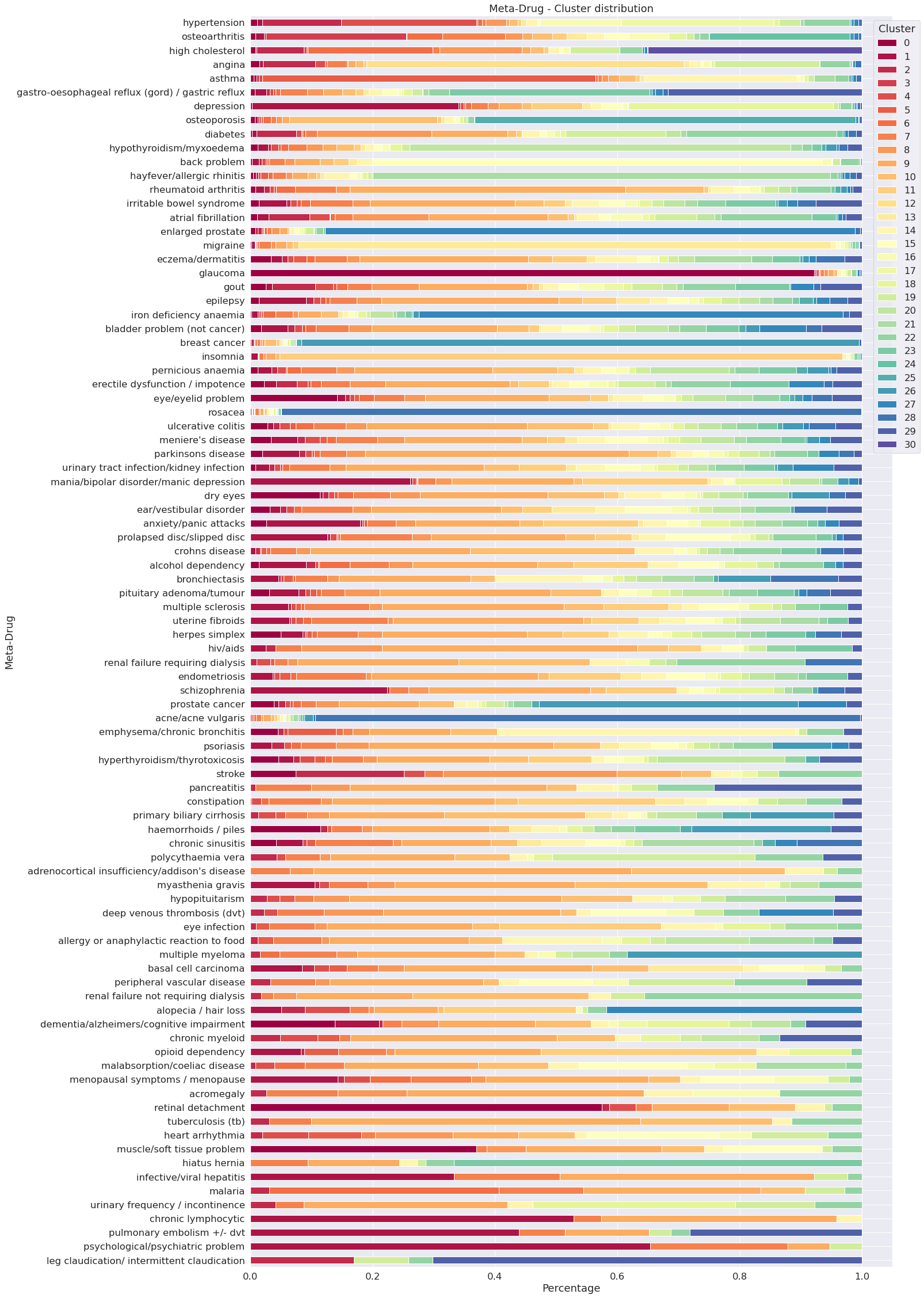}
      \caption{Meta-Drug distributions for the main clusters. The Meta-Drugs are ordered by their frequency among all main clusters.\label{fig:c31-meta-drug-distributions}}
\end{figure}

Figure \ref{fig:c31-meta-drug-distributions} illustrates the distributions of each Meta-Drug, in percentage, among the main clusters. It is noticeable that while a few Meta-Drugs (including "meta-enlarged prostate", "meta-glaucoma", "meta-breast cancer" and "meta-rosacea") tend to have a high percentage representation and are enriched in very few clusters, the majority of the other Meta-Drugs are more evenly distributed, including the Meta-Drugs "meta-osteoarthritis" and "meta-hypertension". 

The Meta-Drug assignment reveals interesting cluster aspects. For example, $C3$ features a single Meta-Drug "meta-osteoarthritis". This indicates that all medications in that cluster are prescribed due to rheumatic symptoms. Interestingly, participants in $C3$ have - next to osteoarthritis - other top diseases, including asthma, hay fever and hypertension, which are not therapeutically addressed via a drug prescription (i.e., no corresponding Meta-Drug), indicating that the severity of these additional diseases is low. In contrast, while the top Meta-Drug in $C7$ is also "meta-osteoarthritis", there are other Meta-Drugs including "meta-gastric reflux", "meta-angina" and "meta-hypothyroidism", also reflected by the top diseases. The presence of additional Meta-Drugs is indicative of active disease needing drug treatment.

\subsubsection*{Assessment of drug-drug interactions and drug metabolizing gene variants in the clusters}

We are particularly interested whether membership in a certain cluster exposes a participant to specific DDIs.  That is, the specific drug combinations in a cluster may give rise to very distinct interactions and side effects.

\begin{figure}[ht]
  \centering
      \includegraphics[width=1.\textwidth,height=\textheight,keepaspectratio]{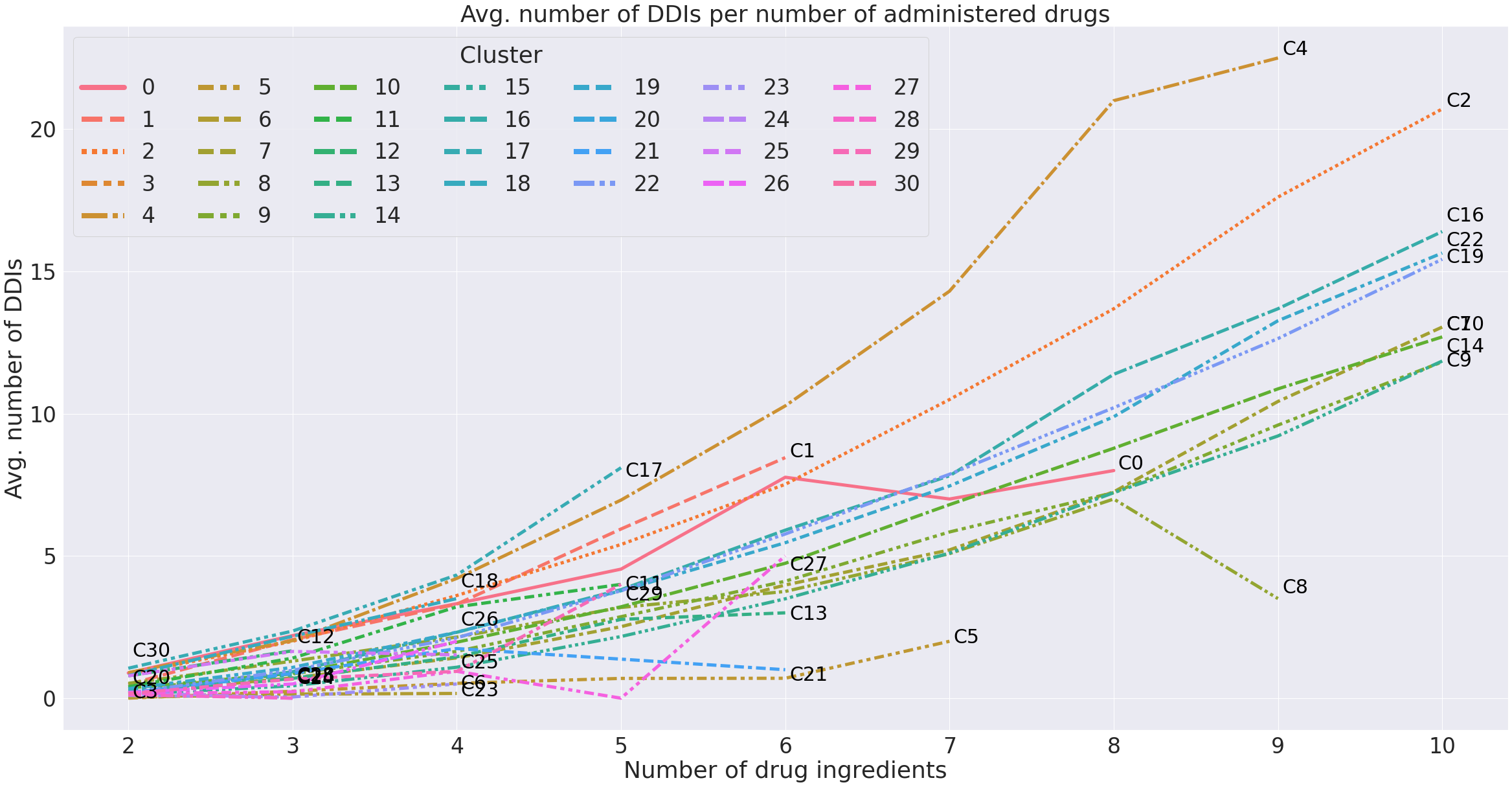}
      \caption{Average number of DDIs per main cluster for participants with up to $10$ drug ingredients, based on all $60$ distinct DDI types. \label{fig:c31-ddiType-avg}}
\end{figure}

\begin{figure}[ht]
  \centering
      \includegraphics[width=1.\textwidth,height=\textheight,keepaspectratio]{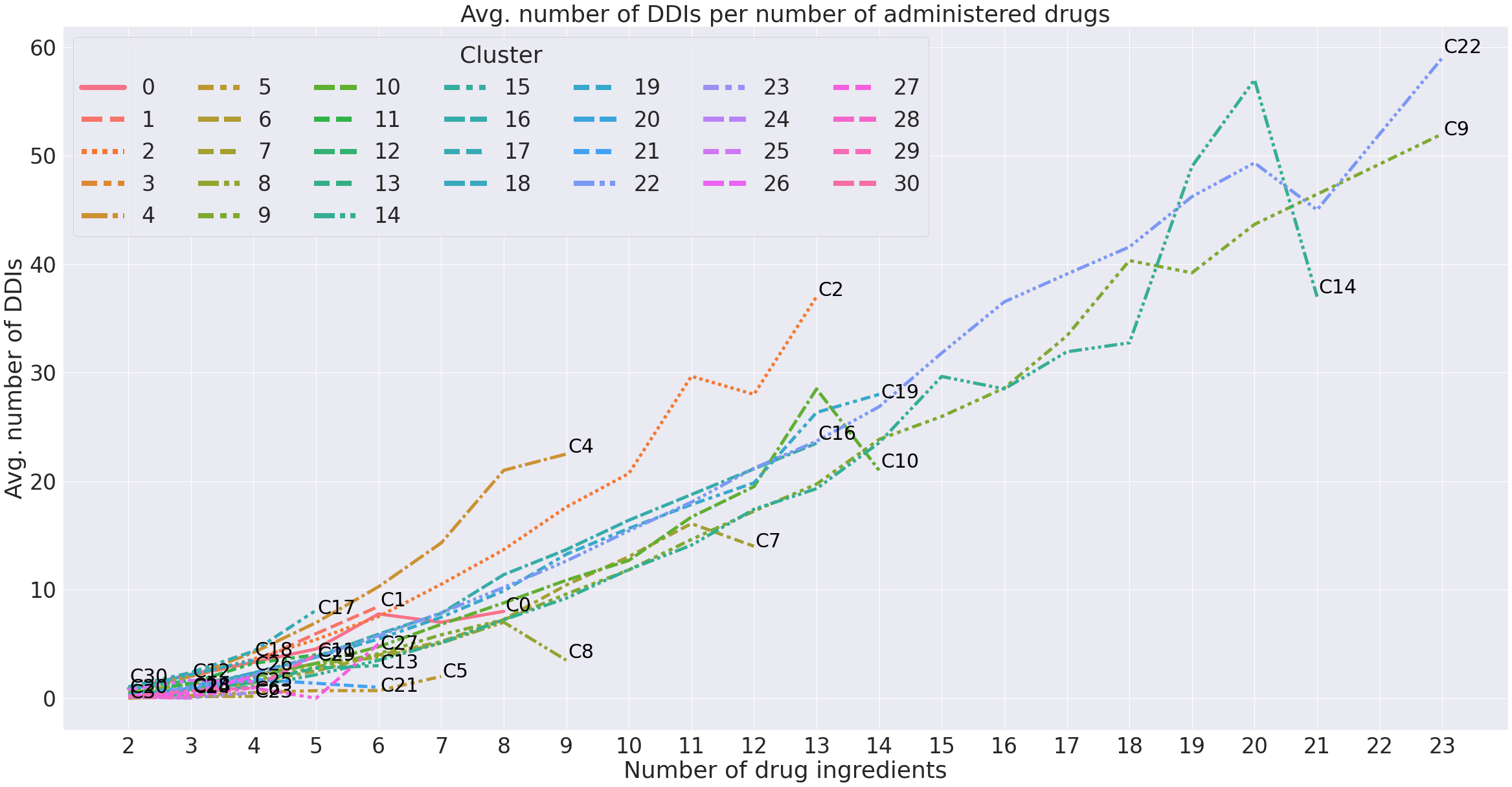}
      \caption{Average number of DDIs per main cluster for participants with up to $23$ drug ingredients, based an all $60$ distinct DDI types. \label{fig:c31-ddiType-avg-23}}
\end{figure}

We therefore investigated pairwise interactions of drugs for each participant and mapped those to a corresponding DDI (see Table \ref{table:ddi_types} and Methods for more details - an example participant with their original drug prescription and DDI mappings is shown in Figure \ref{fig:participant-example}). Notably,  we find cluster-specific DDI patterns. Figure \ref{fig:c31-ddiType-avg} depicts the mean proportions of $60$ types of DDIs for participants taking $1$ to $10$ different drugs. Strikingly, the increase of DDIs per additionally prescribed drug is different for the individual drug constellation clusters (Figure \ref{fig:c31-ddiType-avg-23} shows an expanded graph including participants that take up to $23$ different drugs). Notably, some clusters have consistently a low average number of DDIs, e.g., $C21$ and $C5$, across an increasing number of drugs. In contrast, participants in clusters $C2$ and $C4$ depict a higher rate and number of DDIs. 

\begin{figure}[ht]
  \centering
      \includegraphics[width=\textwidth,height=0.95\textheight,keepaspectratio]{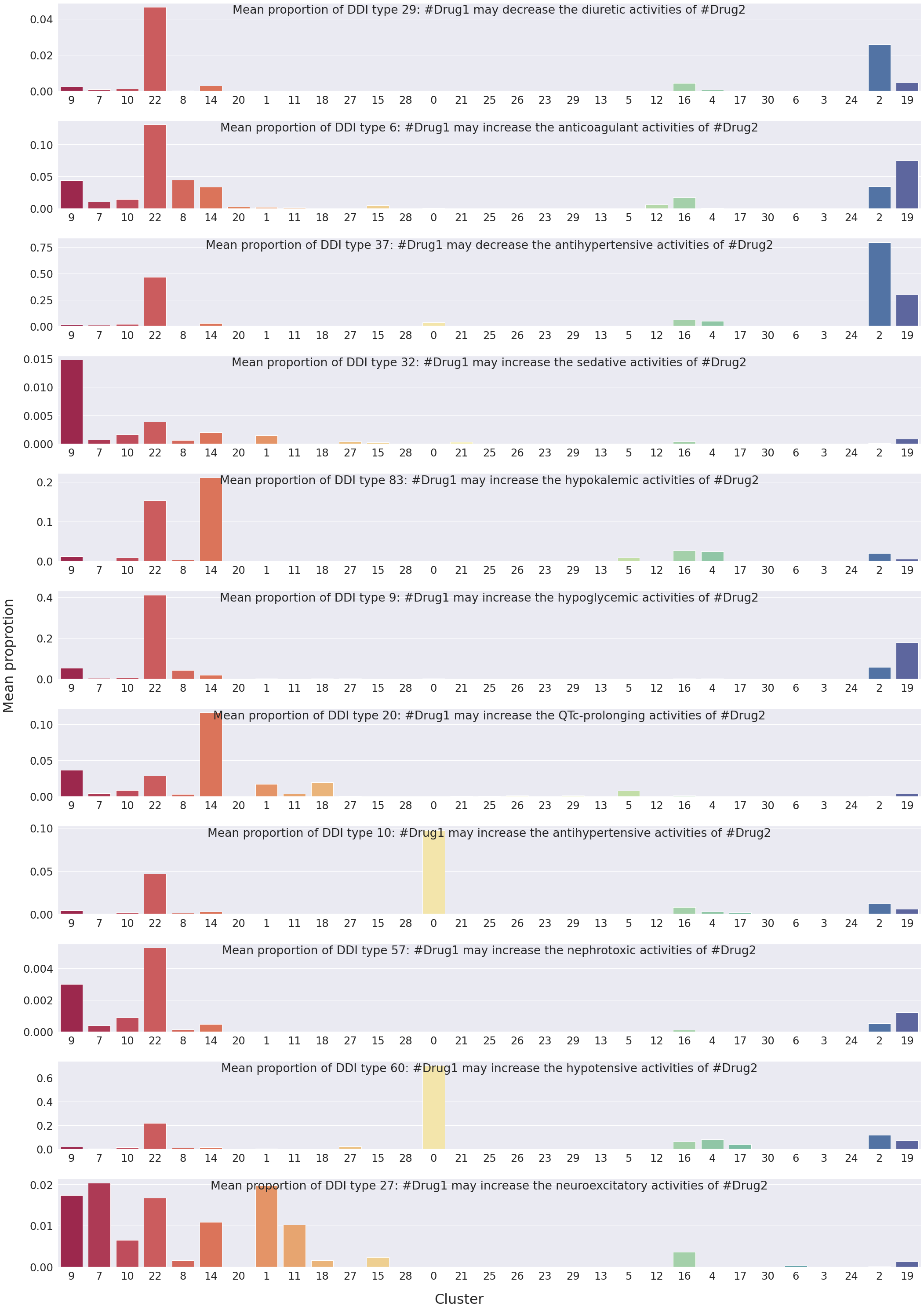}
      \caption{Distributions of $11$ physiologically specific DDI types for the main clusters. The clusters are ordered according to their similarity in terms of Meta-Drug enrichment (Figure \ref{fig:c31-metadrug-enrichment}).\label{fig:c31-ddiType-distributions}}
\end{figure}

Furthermore, Figure \ref{fig:c31-ddiType-distributions} depicts the mean proportions - per cluster - of $11$ selected DDIs with specific physiologic consequences (see table \ref{table:ddi_types}). For example, DDI type $60$ ("Drug A may increase the hypotensive activities of Drug B") is clearly prevalent in $C0$ compared to the other clusters. Glaucoma is the top disease in $C0$ and the combination of the drugs Dorzolamide and Timolol (both used in the treatment of elevated intraocular pressure in ocular hypertension) in $C0$ might lead to a type $60$ DDI, resulting in synergistic blood pressure lowering.

\begin{figure}[ht]
  \centering
      \includegraphics[width=\textwidth,height=\textheight,keepaspectratio]{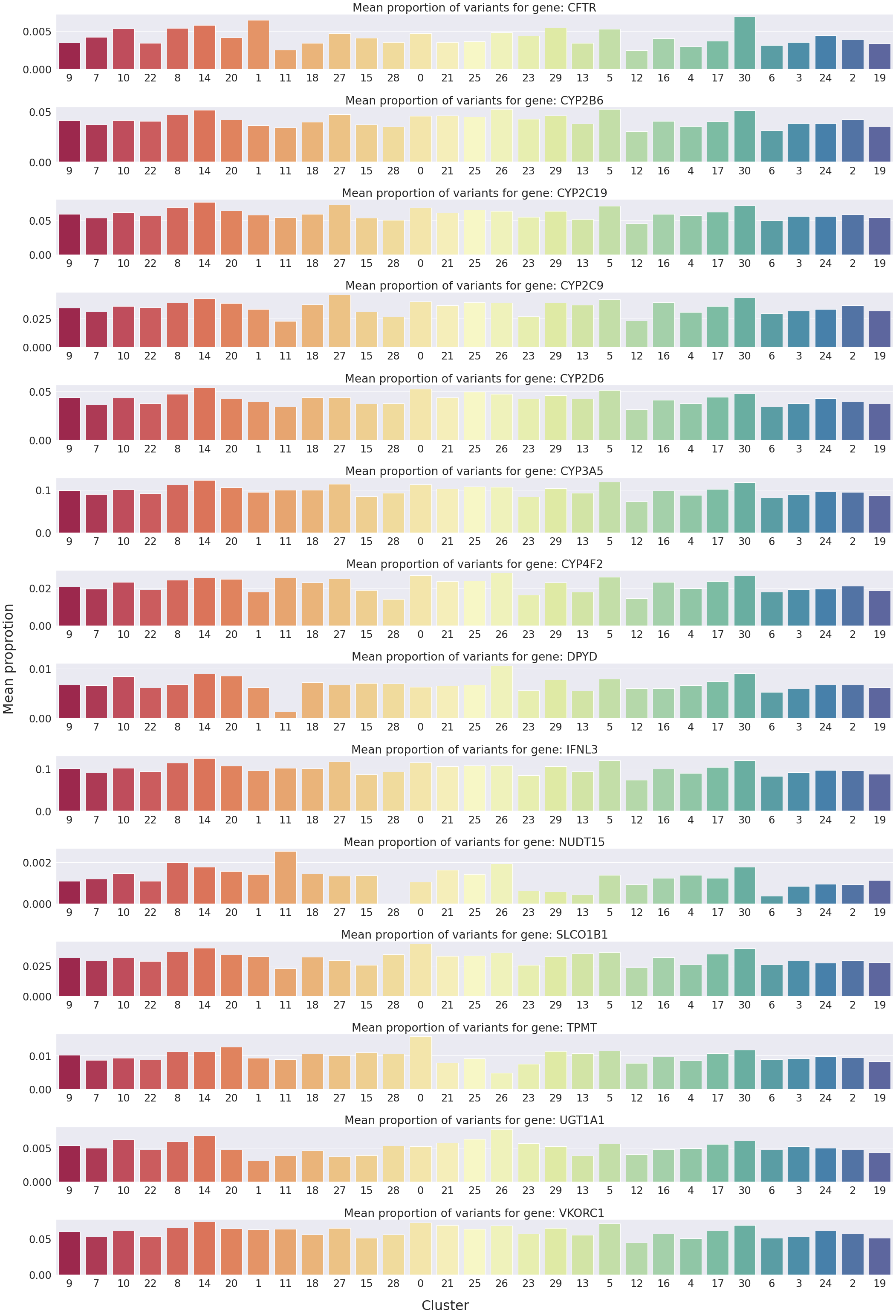}
      \caption{Pharmacogene genetic variant distributions for the main clusters.\label{fig:c31-genetic-variant-distributions}}
\end{figure}

We also investigated the presence of drug metabolizing gene variants (GV) across the clusters (Figure \ref{fig:c31-genetic-variant-distributions}). Our results indicate an even distribution of GVs across the clusters. This is expected, as the clustering was predominantly based on drug prescriptions. 

\clearpage
\subsection*{Comparison of drug constellation clusters}

One of our main motivation is to elucidate common drug prescription patterns and their possible effects on patient health. One obvious approach is to inspect two (or more) clusters, tabulate observed differences in drug combinations and known side effects (DDIs) and link this information with the participants' health outcomes. 

\begin{figure}[ht]
  \centering
      \includegraphics[width=0.66\textwidth,keepaspectratio]{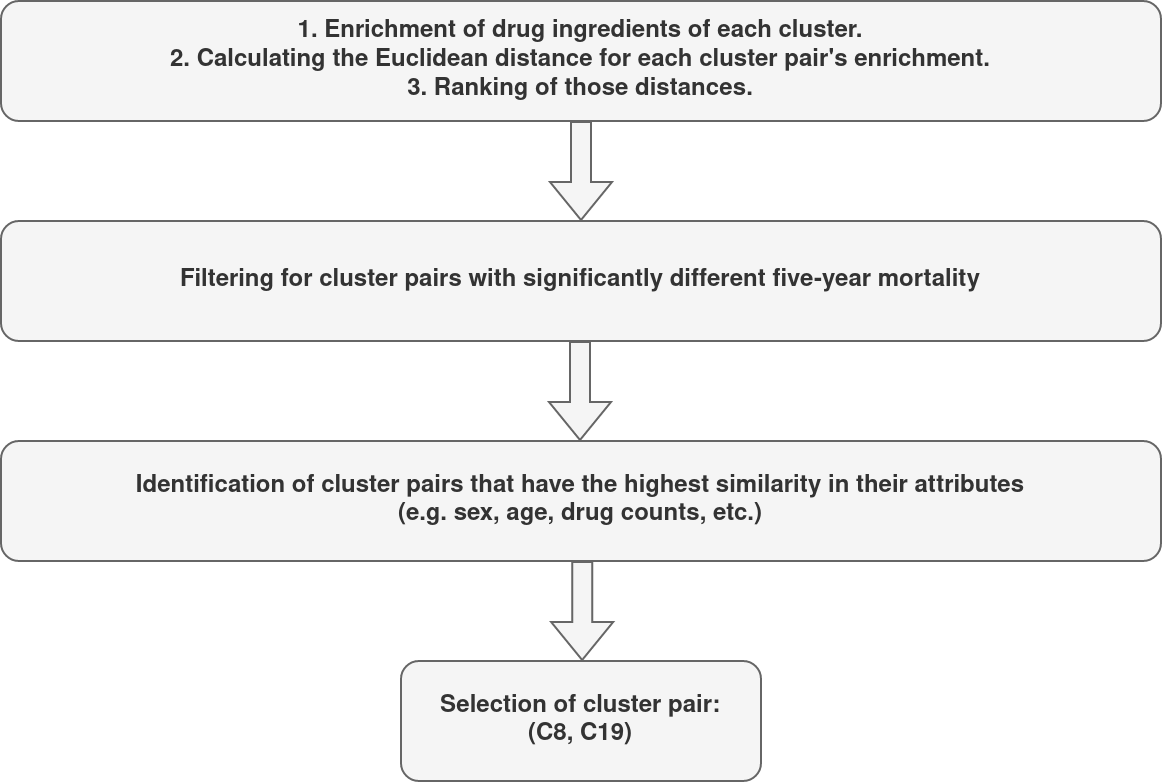}
      \caption{Selection of the cluster pair for the case study.\label{fig:cluster-selection}}
\end{figure}

To demonstrate the opportunities for knowledge discovery from our data set, we compared one pair of clusters with similar (but not identical) drug prescription profiles and health attributes but different health outcomes. The selection process of that pair is depicted in Figure \ref{fig:cluster-selection}. 

\subsubsection*{Comparison of clusters C8 an C19: Exploring the role of DDIs in hypertension control}

\begin{figure}[ht]
  \centering
      \includegraphics[width=\textwidth,height=\textheight,keepaspectratio]{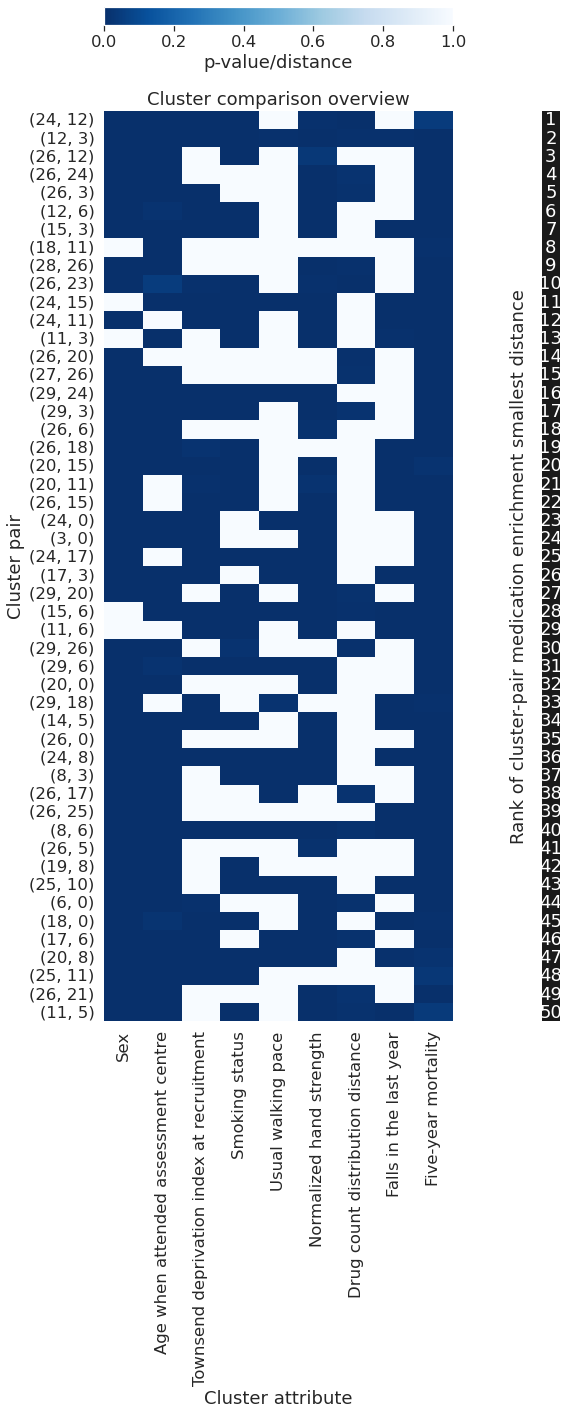}
      \caption{Cluster comparison for cluster pairs with significantly different five-year mortality - basic attributes.\label{fig:c31-cluster-comparison}}
\end{figure}

We performed pairwise comparison of the main clusters in terms of five-year mortality, drug prescription constellations and health attributes according to Figure \ref{fig:cluster-selection}. The procedure allowed us to identify possible cluster candidates for further analysis (see also Figure \ref{fig:c31-cluster-comparison}). For this study, we decided to focus on the comparison of clusters $C8$ and $C19$, with different five-year mortality, but otherwise similar profiles. 

\begin{figure}[ht]
  \centering
      \includegraphics[width=\textwidth,height=\textheight,keepaspectratio]{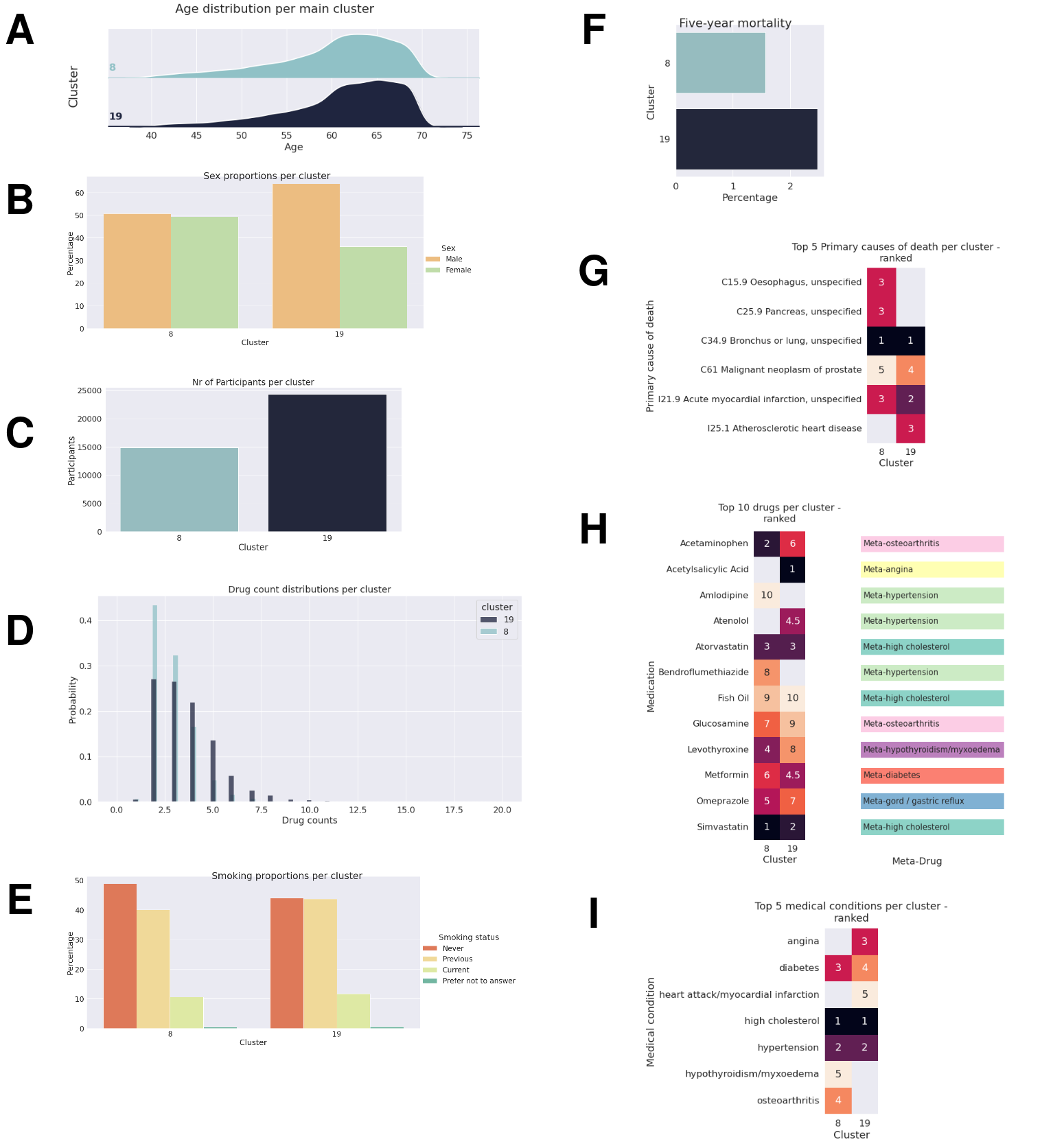}
      \caption{Comparison of clusters $C8$, $C19$. \textbf{A} Age distributions of the two clusters. \textbf{B} Sex proportions. \textbf{C} Number of participants. \textbf{D} Drug count frequencies. \textbf{E} Smoking status proportions. \textbf{F} Five-year mortality. \textbf{G} Top five primary causes of death, ranked. \textbf{H} Top $10$ drugs per cluster, ranked. \textbf{I} Top five medical conditions, ranked. \label{fig:c8c19-overview}}
\end{figure}

Both clusters feature hypertension, high cholesterol and diabetes among the top five diseases. While $C19$ features higher ranks for cardiovascular deaths among the primary causes of death, $C8$ has more  cancer-related (esp. esophageal, pancreatic) deaths (Figure \ref{fig:c8c19-overview} - G). We expected a cluster with a higher proportion of cancer participants to have a higher five-year mortality. However, $C8$ has significantly lower five-year mortality than $C19$ ($1.6 \%$ vs. $2.5\%$, $p = 10^{-6}$, two-proportions Z-test) (Figure \ref{fig:c8c19-overview} - F). Accordingly, we examined which factors of interest, i.e., medications, diseases, DDIs and GVs, could possibly explain the difference in five-year mortality. 

Reasons for higher five-year mortality in $C19$:  Part of the difference in five-year mortality might be explained by remaining discrepancies in the underlying diseases. While high cholesterol, hypertension and diabetes are shared top diseases between the two clusters, $C19$ also lists angina and heart attack/myocardial infarction among its chief disorders. Furthermore, some of the higher five-year mortality of $C19$ compared to $C8$ can be explained by the slightly higher age distribution and more unfavorable smoking status of $C19$ (Figure \ref{fig:c8c19-overview} - A and B). 

Reasons for lower mortality in $C8$: The more favorable five-year mortality outcome of $C8$ might be attributed to differences in medication as well as differences in the distributions of drug metabolizing enzymes and DDIs:

1) Antihypertensive medications. $C8$ includes distinct antihypertensive medications in the top $10$ drugs (Figure \ref{fig:c8c19-overview} - H) compared to $C19$ (i.e., Amlodipine, Bendroflumethiazide in $C8$ vs. Atenolol in $C19$). Randomized controlled trials (RCTs) and meta-analysis have shown that beta blockers (Atenolol) do not improve mortality for participants with hypertension and do not have any protective effect with regard to coronary artery disease in comparison to other antihypertensive medications such as calcium channel blockers (Amlodipine) \cite{ziff2020beta}.

\begin{figure}[ht]
  \centering
      \includegraphics[width=\textwidth,height=\textheight,keepaspectratio]{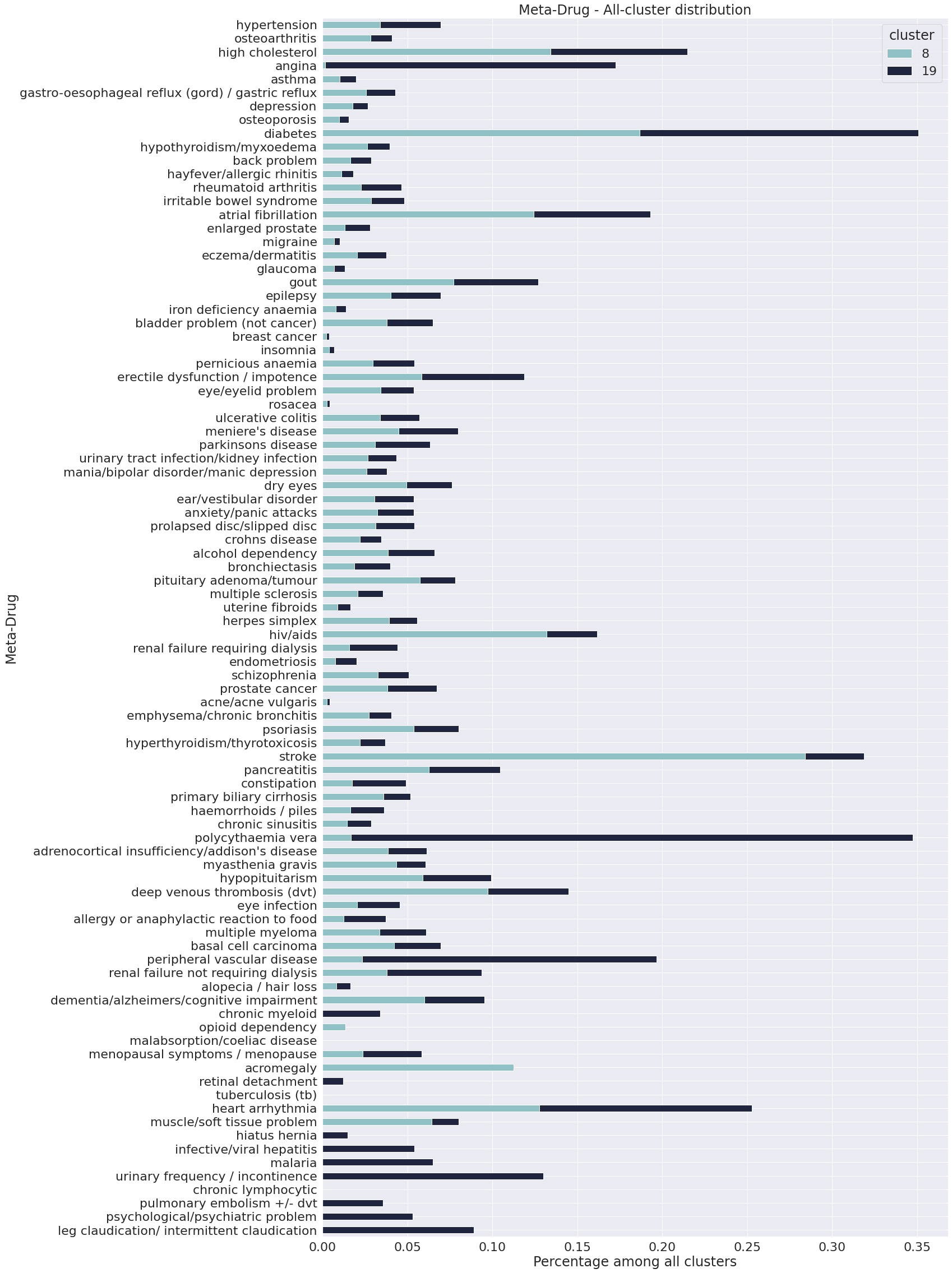}
      \caption{Meta-Drug distributions for cluster pair C8, C19. \label{fig:casestudy-metadrug}}
\end{figure}

2) Lipid-lowering medications. Participants in $C8$ have more lipid-lowering medications (Figure \ref{fig:casestudy-metadrug}) than participants in $C19$. Specifically, Simvastatin has a higher rank in the top $10$ drugs of $C8$ (Figure \ref{fig:c8c19-overview} - H). Lipid-lowering medications, specifically statins, have been shown to reduce major vascular events and vascular mortality \cite{armitage2019efficacy}.

\begin{figure}[ht]
  \centering
      \includegraphics[width=\textwidth,height=\textheight,keepaspectratio]{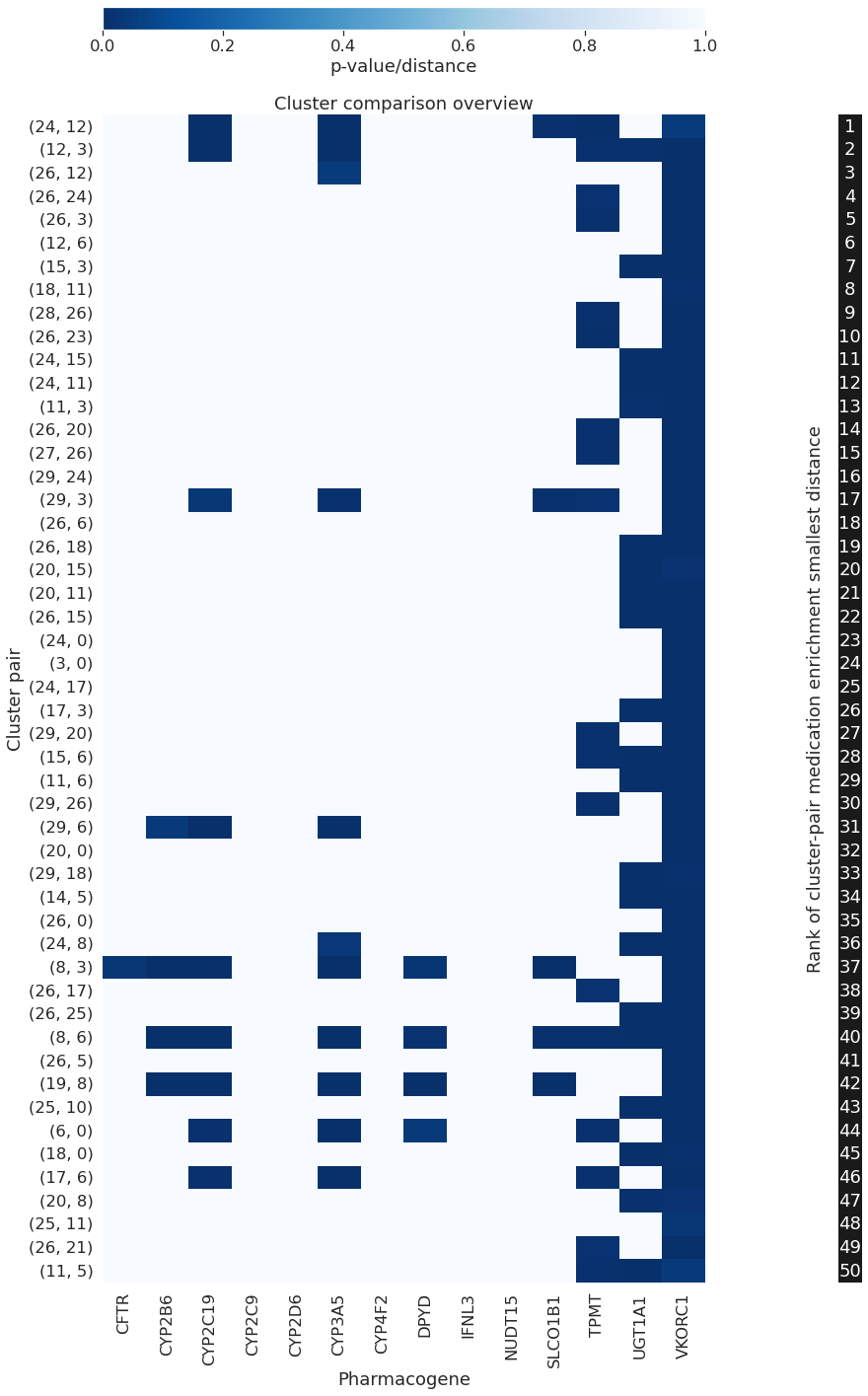}
      \caption{Cluster comparison for cluster pairs with significantly different five-year mortality - pharmacogene variants.\label{fig:c31-cluster-comparison-genes}}
\end{figure}

3) Distribution of drug metabolizing GVs. The two clusters include statistically significantly different proportions ($0.0362$ for $C8$ vs. $0.0277$ for $C19$, $p = 3\cdot10^{-6}$, two-proportions Z-test) of drug metabolizing gene \textit{SLCO1B1} variants (Figure \ref{fig:c31-cluster-comparison-genes}), which have known clinical effect on the metabolism of Simvastatin. GVs of \textit{SLCO1B1}, a hepatic transporter of statins, might lead to a reduction in uptake of statins in the liver, where they are metabolized \cite{noauthor_cpic_nodate}. This alteration can increase the lipid lowering effect of statins leading to a decrease in atherosclerosis. As these variants are more common among participants of $C8$ they might be at a lower risk for atherosclerosis.

4) Differences in DDIs. The DDI type $37$ (i.e., "Drug A may decrease antihypertensive activities of Drug B") is elevated in $C19$ compared to $C8$ (Figure \ref{fig:c31-ddiType-distributions}), which can lead to a decreased antihypertensive effect in $C19$. For example, the drug combinations Acetylsalicylic acid-Atenolol and Salbutamol-Ramipril that are found in $C19$ might result in a type $37$ DDI for some of its participants. To further investigate the last point we split the participants of $C19$ into two groups, according to table \ref{table:comparison_c19}. For each group we only included participants that have at least one of the top five diseases in this cluster (Figure \ref{fig:c8c19-overview} - I). As expected, participants with no type $37$ DDIs (decreased antihypertensive side effect) have a lower five-year mortality ($p=0.033$, two-proportions Z-test) compared to participants that have at least one of those deleterious interactions. These findings suggests that non-optimal drug combinations in $C19$ could negatively affect participant outcomes.  Importantly, part of this discrepancy might also be attributed to the different confounding factors (e.g., age or sex distributions between these groups).

\begin{table}[ht]
\begin{center}
    \normalsize
\begin{tabular}{l|r|r|r|r}
\textbf{Group} & \textbf{$\#$ Participants} & \textbf{$\%$ Female} & \textbf{Avg. age} & \textbf{$\%$ Five-year mortality}
\\\hline
No type $37$ DDIs & $12'595$ & $36$ & $61.7$ & $2.3$ \\
$1+$ type $37$ DDIs & $6'303$ & $29$ & $61.9$ & $2.8$  \\

\end{tabular}
    \end{center}
    \caption{Comparison of participants with and without DDI type $37$ within $C19$.\label{table:comparison_c19}}
\end{table}

\clearpage

\section*{Discussion}

\subsection*{Pharmacological overview of the UK Biobank}

The UK Biobank, with over $500'000$ participants, is a rich biomedical dataset that allows for unprecedented studies into care practices, disease etiology and prevention. Here, we provide an overview of the drug prescription landscape of UK Biobank participants with the aim of identifying groups of participants sharing similar drug constellations. To that end, we introduce a novel method for comparing drug prescriptions between participants, involving data-driven learning of \textbf{Meta-Drugs}, i.e., groups of drugs that treat similar diseases and reflect drug prescriptions by indication, rather than active ingredient.  \textbf{Meta-Drugs} allow us to group UK Biobank participants and to derive drug constellations reflecting both the active ingredients as well as the underlying diseases and drug indications. 

We were particularly interested in studying the relationships between polypharmacy, drug constellations and  drug-drug interactions that occur as a result of concurrent drug administration. Polypharmacy traditionally investigates patient outcome as a function of the number of drugs prescribed. Here, we additionally study outcome with regard to the specific prescribed drug combinations. 

Our study revealed $31$ unique drug constellations in the UK Biobank, mostly driven by differences in the \textbf{Meta-Drug} distributions. We show that \textbf{drug constellations matter in terms of expected DDIs, both in terms of the types and number of DDIs}. We then investigated whether differences in drug constellations and DDIs may explain differences in participant outcomes. We present a case study of two clusters with highly similar participants in terms of underlying diseases, age and sex distribution, but different five-year mortality. Particularly, both clusters contained a high number of participants suffering from hypertension that are treated with antihypertensive medication. Investigating the differences between these two drug prescription clusters, we found differences in the antihypertensive treatment, as well as the frequency of a DDI that may lead to a decreased antihypertensive drug effect. Other possible causes of the observed difference in five-year mortality include differences associated with lipid-lowering drug medications, including frequency of prescription and presence of GVs affecting their metabolism. Altogether, we believe that our approach is a suitable vehicle to derive testable hypotheses of the effect of drug combinations and DDIs on patient outcomes.

\subsection*{Limitations}

While the purpose of this study was not to uncover causal effects between participants characteristics and their outcomes, we would like to point out that future studies might improve on our analysis by introducing additional confounder control strategies, beyond the statistical comparison of selected variables used in this study. This may include the use of propensity score-like approaches to balance the distribution of participant characteristics before testing the effect of specific variables (like DDIs) on participant outcomes.  

\subsection*{Conclusions}

This study provides an overview of the UK Biobank pharmacological landscape, revealing distinct drug constellation patterns driven by the participants' disease status. We show that these constellations matter in terms of the number and types of expected DDIs and are thus an important consideration when studying polypharmacy-related adverse outcomes. 

\clearpage


\subsection*{Availability of data and materials}
The controlled UK Biobank data used in the study is available for investigators submitting a project proposal via the UK Biobank Access Management System. The other data used in this study is publicly available as described in the Methods section.

\subsection*{Competing interests}
    The authors declare that they have no competing interests.

\subsection*{Acknowledgements}
This research has been conducted using the UK Biobank Resource under Application Number 54717.

We especially thank Nicolas Andres Perez Gonzalez for helping with analysis ideas during the initial phase of this study.

\subsection*{Author's contributions}
KS and MK worked on project development and experimental design. KS worked on data processing, analysis workflow, and algorithm and model implementation. KS and DT analyzed and interpreted the data. KS drafted the manuscript. DT, GR and MK supervised and edited the manuscript. GR processed medical information. All authors approved the final article.

\section*{Analysis software}
\begin{tabular}{ l | r | r }
 \textbf{Software} & \textbf{Version} & \textbf{Reference} \\ \hline
 python & $3.8.12$ & \\
 scikit-learn & $0.24.2$ & \\
 scipy & $1.7.1$ & \\
 statsmodels & $0.12.2$ & \cite{seabold2010statsmodels} \\
 rapids & $21.10$ & \cite{rapids2018}\\
 ukbREST & & \cite{pividori2019ukbrest} \\
\end{tabular}


\bibliographystyle{bmc-mathphys} 
\bibliography{article.bib}      


\begin{thebibliography}{20}
\ifx \bisbn   \undefined \def \bisbn  #1{ISBN #1}\fi
\ifx \binits  \undefined \def \binits#1{#1}\fi
\ifx \bauthor  \undefined \def \bauthor#1{#1}\fi
\ifx \batitle  \undefined \def \batitle#1{#1}\fi
\ifx \bjtitle  \undefined \def \bjtitle#1{#1}\fi
\ifx \bvolume  \undefined \def \bvolume#1{\textbf{#1}}\fi
\ifx \byear  \undefined \def \byear#1{#1}\fi
\ifx \bissue  \undefined \def \bissue#1{#1}\fi
\ifx \bfpage  \undefined \def \bfpage#1{#1}\fi
\ifx \blpage  \undefined \def \blpage #1{#1}\fi
\ifx \burl  \undefined \def \burl#1{\textsf{#1}}\fi
\ifx \doiurl  \undefined \def \doiurl#1{\textsf{#1}}\fi
\ifx \betal  \undefined \def \betal{\textit{et al.}}\fi
\ifx \binstitute  \undefined \def \binstitute#1{#1}\fi
\ifx \binstitutionaled  \undefined \def \binstitutionaled#1{#1}\fi
\ifx \bctitle  \undefined \def \bctitle#1{#1}\fi
\ifx \beditor  \undefined \def \beditor#1{#1}\fi
\ifx \bpublisher  \undefined \def \bpublisher#1{#1}\fi
\ifx \bbtitle  \undefined \def \bbtitle#1{#1}\fi
\ifx \bedition  \undefined \def \bedition#1{#1}\fi
\ifx \bseriesno  \undefined \def \bseriesno#1{#1}\fi
\ifx \blocation  \undefined \def \blocation#1{#1}\fi
\ifx \bsertitle  \undefined \def \bsertitle#1{#1}\fi
\ifx \bsnm \undefined \def \bsnm#1{#1}\fi
\ifx \bsuffix \undefined \def \bsuffix#1{#1}\fi
\ifx \bparticle \undefined \def \bparticle#1{#1}\fi
\ifx \barticle \undefined \def \barticle#1{#1}\fi
\ifx \bconfdate \undefined \def \bconfdate #1{#1}\fi
\ifx \botherref \undefined \def \botherref #1{#1}\fi
\ifx \url \undefined \def \url#1{\textsf{#1}}\fi
\ifx \bchapter \undefined \def \bchapter#1{#1}\fi
\ifx \bbook \undefined \def \bbook#1{#1}\fi
\ifx \bcomment \undefined \def \bcomment#1{#1}\fi
\ifx \oauthor \undefined \def \oauthor#1{#1}\fi
\ifx \citeauthoryear \undefined \def \citeauthoryear#1{#1}\fi
\ifx \endbibitem  \undefined \def \endbibitem {}\fi
\ifx \bconflocation  \undefined \def \bconflocation#1{#1}\fi
\ifx \arxivurl  \undefined \def \arxivurl#1{\textsf{#1}}\fi
\csname PreBibitemsHook\endcsname

\bibitem{kantor_trends_2015}
\begin{barticle}
\bauthor{\bsnm{Kantor}, \binits{E.D.}},
\bauthor{\bsnm{Rehm}, \binits{C.D.}},
\bauthor{\bsnm{Haas}, \binits{J.S.}},
\bauthor{\bsnm{Chan}, \binits{A.T.}},
\bauthor{\bsnm{Giovannucci}, \binits{E.L.}}:
\batitle{Trends in {Prescription} {Drug} {Use} {Among} {Adults} in the {United}
  {States} {From} 1999-2012}.
\bjtitle{JAMA}
\bvolume{314}(\bissue{17}),
\bfpage{1818}--\blpage{1830}
(\byear{2015}).
doi:\doiurl{10.1001/jama.2015.13766}.
Accessed 2021-11-06
\end{barticle}
\endbibitem

\bibitem{hilmer_effects_2009}
\begin{barticle}
\bauthor{\bsnm{Hilmer}, \binits{S.}},
\bauthor{\bsnm{Gnjidic}, \binits{D.}}:
\batitle{The {Effects} of {Polypharmacy} in {Older} {Adults}}.
\bjtitle{Clinical Pharmacology \& Therapeutics}
\bvolume{85}(\bissue{1}),
\bfpage{86}--\blpage{88}
(\byear{2009}).
doi:\doiurl{10.1038/clpt.2008.224}.
\bcomment{\_eprint:
  https://onlinelibrary.wiley.com/doi/pdf/10.1038/clpt.2008.224}.
Accessed 2021-11-06
\end{barticle}
\endbibitem

\bibitem{maher2014clinical}
\begin{barticle}
\bauthor{\bsnm{Maher}, \binits{R.L.}},
\bauthor{\bsnm{Hanlon}, \binits{J.}},
\bauthor{\bsnm{Hajjar}, \binits{E.R.}}:
\batitle{Clinical consequences of polypharmacy in elderly}.
\bjtitle{Expert opinion on drug safety}
\bvolume{13}(\bissue{1}),
\bfpage{57}--\blpage{65}
(\byear{2014})
\end{barticle}
\endbibitem

\bibitem{sudlow_uk_2015}
\begin{barticle}
\bauthor{\bsnm{Sudlow}, \binits{C.}},
\bauthor{\bsnm{Gallacher}, \binits{J.}},
\bauthor{\bsnm{Allen}, \binits{N.}},
\bauthor{\bsnm{Beral}, \binits{V.}},
\bauthor{\bsnm{Burton}, \binits{P.}},
\bauthor{\bsnm{Danesh}, \binits{J.}},
\bauthor{\bsnm{Downey}, \binits{P.}},
\bauthor{\bsnm{Elliott}, \binits{P.}},
\bauthor{\bsnm{Green}, \binits{J.}},
\bauthor{\bsnm{Landray}, \binits{M.}},
\bauthor{\bsnm{Liu}, \binits{B.}},
\bauthor{\bsnm{Matthews}, \binits{P.}},
\bauthor{\bsnm{Ong}, \binits{G.}},
\bauthor{\bsnm{Pell}, \binits{J.}},
\bauthor{\bsnm{Silman}, \binits{A.}},
\bauthor{\bsnm{Young}, \binits{A.}},
\bauthor{\bsnm{Sprosen}, \binits{T.}},
\bauthor{\bsnm{Peakman}, \binits{T.}},
\bauthor{\bsnm{Collins}, \binits{R.}}:
\batitle{{UK} {Biobank}: {An} {Open} {Access} {Resource} for {Identifying} the
  {Causes} of a {Wide} {Range} of {Complex} {Diseases} of {Middle} and {Old}
  {Age}}.
\bjtitle{PLOS Medicine}
\bvolume{12}(\bissue{3}),
\bfpage{1001779}
(\byear{2015}).
doi:\doiurl{10.1371/journal.pmed.1001779}.
\bcomment{Publisher: Public Library of Science}.
Accessed 2021-11-06
\end{barticle}
\endbibitem

\bibitem{anastopoulos_multi-drug_2021}
\begin{barticle}
\bauthor{\bsnm{Anastopoulos}, \binits{I.N.}},
\bauthor{\bsnm{Herczeg}, \binits{C.K.}},
\bauthor{\bsnm{Davis}, \binits{K.N.}},
\bauthor{\bsnm{Dixit}, \binits{A.C.}}:
\batitle{Multi-{Drug} {Featurization} and {Deep} {Learning} {Improve}
  {Patient}-{Specific} {Predictions} of {Adverse} {Events}}.
\bjtitle{International Journal of Environmental Research and Public Health}
\bvolume{18}(\bissue{5}),
\bfpage{2600}
(\byear{2021}).
doi:\doiurl{10.3390/ijerph18052600}.
\bcomment{Number: 5 Publisher: Multidisciplinary Digital Publishing Institute}.
Accessed 2021-12-17
\end{barticle}
\endbibitem

\bibitem{zemedikun_patterns_2018}
\begin{barticle}
\bauthor{\bsnm{Zemedikun}, \binits{D.T.}},
\bauthor{\bsnm{Gray}, \binits{L.J.}},
\bauthor{\bsnm{Khunti}, \binits{K.}},
\bauthor{\bsnm{Davies}, \binits{M.J.}},
\bauthor{\bsnm{Dhalwani}, \binits{N.N.}}:
\batitle{Patterns of {Multimorbidity} in {Middle}-{Aged} and {Older} {Adults}:
  {An} {Analysis} of the {UK} {Biobank} {Data}}.
\bjtitle{Mayo Clinic Proceedings}
\bvolume{93}(\bissue{7}),
\bfpage{857}--\blpage{866}
(\byear{2018}).
doi:\doiurl{10.1016/j.mayocp.2018.02.012}.
Accessed 2021-12-17
\end{barticle}
\endbibitem

\bibitem{mcqueenie_multimorbidity_2020}
\begin{barticle}
\bauthor{\bsnm{McQueenie}, \binits{R.}},
\bauthor{\bsnm{Foster}, \binits{H.M.E.}},
\bauthor{\bsnm{Jani}, \binits{B.D.}},
\bauthor{\bsnm{Katikireddi}, \binits{S.V.}},
\bauthor{\bsnm{Sattar}, \binits{N.}},
\bauthor{\bsnm{Pell}, \binits{J.P.}},
\bauthor{\bsnm{Ho}, \binits{F.K.}},
\bauthor{\bsnm{Niedzwiedz}, \binits{C.L.}},
\bauthor{\bsnm{Hastie}, \binits{C.E.}},
\bauthor{\bsnm{Anderson}, \binits{J.}},
\bauthor{\bsnm{Mark}, \binits{P.B.}},
\bauthor{\bsnm{Sullivan}, \binits{M.}},
\bauthor{\bsnm{O’Donnell}, \binits{C.A.}},
\bauthor{\bsnm{Mair}, \binits{F.S.}},
\bauthor{\bsnm{Nicholl}, \binits{B.I.}}:
\batitle{Multimorbidity, polypharmacy, and {COVID}-19 infection within the {UK}
  {Biobank} cohort}.
\bjtitle{PLOS ONE}
\bvolume{15}(\bissue{8}),
\bfpage{0238091}
(\byear{2020}).
doi:\doiurl{10.1371/journal.pone.0238091}.
\bcomment{Publisher: Public Library of Science}.
Accessed 2021-12-17
\end{barticle}
\endbibitem

\bibitem{wu_genome-wide_2019}
\begin{barticle}
\bauthor{\bsnm{Wu}, \binits{Y.}},
\bauthor{\bsnm{Byrne}, \binits{E.M.}},
\bauthor{\bsnm{Zheng}, \binits{Z.}},
\bauthor{\bsnm{Kemper}, \binits{K.E.}},
\bauthor{\bsnm{Yengo}, \binits{L.}},
\bauthor{\bsnm{Mallett}, \binits{A.J.}},
\bauthor{\bsnm{Yang}, \binits{J.}},
\bauthor{\bsnm{Visscher}, \binits{P.M.}},
\bauthor{\bsnm{Wray}, \binits{N.R.}}:
\batitle{Genome-wide association study of medication-use and associated disease
  in the {UK} {Biobank}}.
\bjtitle{Nature Communications}
\bvolume{10}(\bissue{1}),
\bfpage{1891}
(\byear{2019}).
doi:\doiurl{10.1038/s41467-019-09572-5}.
\bcomment{Bandiera\_abtest: a Cc\_license\_type: cc\_by Cg\_type: Nature
  Research Journals Number: 1 Primary\_atype: Research Publisher: Nature
  Publishing Group Subject\_term: Genetics research;Genome-wide association
  studies;Genotype Subject\_term\_id:
  genetics-research;genome-wide-association-studies;genotype}.
Accessed 2021-12-09
\end{barticle}
\endbibitem

\bibitem{mcinnes_pharmacogenetics_2021}
\begin{barticle}
\bauthor{\bsnm{McInnes}, \binits{G.}},
\bauthor{\bsnm{Lavertu}, \binits{A.}},
\bauthor{\bsnm{Sangkuhl}, \binits{K.}},
\bauthor{\bsnm{Klein}, \binits{T.E.}},
\bauthor{\bsnm{Whirl-Carrillo}, \binits{M.}},
\bauthor{\bsnm{Altman}, \binits{R.B.}}:
\batitle{Pharmacogenetics at {Scale}: {An} {Analysis} of the {UK} {Biobank}}.
\bjtitle{Clinical Pharmacology \& Therapeutics}
\bvolume{109}(\bissue{6}),
\bfpage{1528}--\blpage{1537}
(\byear{2021}).
doi:\doiurl{10.1002/cpt.2122}.
\bcomment{\_eprint: https://onlinelibrary.wiley.com/doi/pdf/10.1002/cpt.2122}.
Accessed 2021-11-06
\end{barticle}
\endbibitem

\bibitem{Huang2021tdc}
\begin{botherref}
\oauthor{\bsnm{Huang}, \binits{K.}},
\oauthor{\bsnm{Fu}, \binits{T.}},
\oauthor{\bsnm{Gao}, \binits{W.}},
\oauthor{\bsnm{Zhao}, \binits{Y.}},
\oauthor{\bsnm{Roohani}, \binits{Y.}},
\oauthor{\bsnm{Leskovec}, \binits{J.}},
\oauthor{\bsnm{Coley}, \binits{C.W.}},
\oauthor{\bsnm{Xiao}, \binits{C.}},
\oauthor{\bsnm{Sun}, \binits{J.}},
\oauthor{\bsnm{Zitnik}, \binits{M.}}:
Therapeutics data commons: Machine learning datasets and tasks for drug
  discovery and development.
Proceedings of Neural Information Processing Systems, NeurIPS Datasets and
  Benchmarks
(2021)
\end{botherref}
\endbibitem

\bibitem{vanderpump1995incidence}
\begin{barticle}
\bauthor{\bsnm{Vanderpump}, \binits{M.}},
\bauthor{\bsnm{Tunbrldge}, \binits{W.}},
\bauthor{\bsnm{French}, \binits{J.a.}},
\bauthor{\bsnm{Appleton}, \binits{D.}},
\bauthor{\bsnm{Bates}, \binits{D.}},
\bauthor{\bsnm{Clark}, \binits{F.}},
\bauthor{\bsnm{Evans}, \binits{J.G.}},
\bauthor{\bsnm{Hasan}, \binits{D.}},
\bauthor{\bsnm{Rodgers}, \binits{H.}},
\bauthor{\bsnm{Tunbridge}, \binits{F.}}, \betal:
\batitle{The incidence of thyroid disorders in the community: a twenty-year
  follow-up of the whickham survey}.
\bjtitle{Clinical endocrinology}
\bvolume{43}(\bissue{1}),
\bfpage{55}--\blpage{68}
(\byear{1995})
\end{barticle}
\endbibitem

\bibitem{ziff2020beta}
\begin{barticle}
\bauthor{\bsnm{Ziff}, \binits{O.J.}},
\bauthor{\bsnm{Samra}, \binits{M.}},
\bauthor{\bsnm{Howard}, \binits{J.P.}},
\bauthor{\bsnm{Bromage}, \binits{D.I.}},
\bauthor{\bsnm{Ruschitzka}, \binits{F.}},
\bauthor{\bsnm{Francis}, \binits{D.P.}},
\bauthor{\bsnm{Kotecha}, \binits{D.}}:
\batitle{Beta-blocker efficacy across different cardiovascular indications: an
  umbrella review and meta-analytic assessment}.
\bjtitle{BMC medicine}
\bvolume{18}(\bissue{1}),
\bfpage{1}--\blpage{11}
(\byear{2020})
\end{barticle}
\endbibitem

\bibitem{armitage2019efficacy}
\begin{barticle}
\bauthor{\bsnm{Armitage}, \binits{J.}},
\bauthor{\bsnm{Baigent}, \binits{C.}},
\bauthor{\bsnm{Barnes}, \binits{E.}},
\bauthor{\bsnm{Betteridge}, \binits{D.J.}},
\bauthor{\bsnm{Blackwell}, \binits{L.}},
\bauthor{\bsnm{Blazing}, \binits{M.}},
\bauthor{\bsnm{Bowman}, \binits{L.}},
\bauthor{\bsnm{Braunwald}, \binits{E.}},
\bauthor{\bsnm{Byington}, \binits{R.}},
\bauthor{\bsnm{Cannon}, \binits{C.}}, \betal:
\batitle{Efficacy and safety of statin therapy in older people: a meta-analysis
  of individual participant data from 28 randomised controlled trials}.
\bjtitle{The Lancet}
\bvolume{393}(\bissue{10170}),
\bfpage{407}--\blpage{415}
(\byear{2019})
\end{barticle}
\endbibitem

\bibitem{noauthor_cpic_nodate}
\begin{botherref}
{CPIC}® {Guideline} for {Simvastatin} and {SLCO1B1}.
\url{https://cpicpgx.org/guidelines/guideline-for-simvastatin-and-slco1b1/}
Accessed 2021-12-20
\end{botherref}
\endbibitem

\bibitem{seabold2010statsmodels}
\begin{bchapter}
\bauthor{\bsnm{Seabold}, \binits{S.}},
\bauthor{\bsnm{Perktold}, \binits{J.}}:
\bctitle{statsmodels: Econometric and statistical modeling with python}.
In: \bbtitle{9th Python in Science Conference}
(\byear{2010})
\end{bchapter}
\endbibitem

\bibitem{rapids2018}
\begin{botherref}
\oauthor{\bsnm{Team}, \binits{R.D.}}:
RAPIDS: Collection of Libraries for End to End GPU Data Science.
(2018).
\url{https://rapids.ai}
\end{botherref}
\endbibitem

\bibitem{pividori2019ukbrest}
\begin{barticle}
\bauthor{\bsnm{Pividori}, \binits{M.}},
\bauthor{\bsnm{Im}, \binits{H.K.}}:
\batitle{ukbrest: efficient and streamlined data access for reproducible
  research in large biobanks}.
\bjtitle{Bioinformatics}
\bvolume{35}(\bissue{11}),
\bfpage{1971}--\blpage{1973}
(\byear{2019})
\end{barticle}
\endbibitem

\bibitem{fisher1935statistical}
\begin{barticle}
\bauthor{\bsnm{Fisher}, \binits{R.A.}}:
\batitle{Statistical methods for research workers}.
\bjtitle{Edinburgh: Oliver and Boyd, 1934 and The logic of inductive interence;
  Royal Statistical Society}
\bvolume{98},
\bfpage{39}
(\byear{1935})
\end{barticle}
\endbibitem

\bibitem{kitchens2002basic}
\begin{botherref}
\oauthor{\bsnm{Kitchens}, \binits{L.}}:
Basic statistics and data analysis. 2003.
Pacific Grove, CA: Thomson Brooks/Cole
(2002)
\end{botherref}
\endbibitem

\bibitem{kantorovich1960mathematical}
\begin{barticle}
\bauthor{\bsnm{Kantorovich}, \binits{L.V.}}:
\batitle{Mathematical methods of organizing and planning production}.
\bjtitle{Management science}
\bvolume{6}(\bissue{4}),
\bfpage{366}--\blpage{422}
(\byear{1960})
\end{barticle}
\endbibitem

\end{thebibliography}

\newcommand{\BMCxmlcomment}[1]{}

\BMCxmlcomment{

<refgrp>

<bibl id="B1">
  <title><p>Trends in {Prescription} {Drug} {Use} {Among} {Adults} in the
  {United} {States} {From} 1999-2012</p></title>
  <aug>
    <au><snm>Kantor</snm><fnm>ED</fnm></au>
    <au><snm>Rehm</snm><fnm>CD</fnm></au>
    <au><snm>Haas</snm><fnm>JS</fnm></au>
    <au><snm>Chan</snm><fnm>AT</fnm></au>
    <au><snm>Giovannucci</snm><fnm>EL</fnm></au>
  </aug>
  <source>JAMA</source>
  <pubdate>2015</pubdate>
  <volume>314</volume>
  <issue>17</issue>
  <fpage>1818</fpage>
  <lpage>-1830</lpage>
  <url>https://doi.org/10.1001/jama.2015.13766</url>
</bibl>

<bibl id="B2">
  <title><p>The {Effects} of {Polypharmacy} in {Older} {Adults}</p></title>
  <aug>
    <au><snm>Hilmer</snm><fnm>S</fnm></au>
    <au><snm>Gnjidic</snm><fnm>D</fnm></au>
  </aug>
  <source>Clinical Pharmacology \& Therapeutics</source>
  <pubdate>2009</pubdate>
  <volume>85</volume>
  <issue>1</issue>
  <fpage>86</fpage>
  <lpage>-88</lpage>
  <url>https://onlinelibrary.wiley.com/doi/abs/10.1038/clpt.2008.224</url>
  <note>\_eprint:
  https://onlinelibrary.wiley.com/doi/pdf/10.1038/clpt.2008.224</note>
</bibl>

<bibl id="B3">
  <title><p>Clinical consequences of polypharmacy in elderly</p></title>
  <aug>
    <au><snm>Maher</snm><fnm>RL</fnm></au>
    <au><snm>Hanlon</snm><fnm>J</fnm></au>
    <au><snm>Hajjar</snm><fnm>ER</fnm></au>
  </aug>
  <source>Expert opinion on drug safety</source>
  <publisher>Taylor \& Francis</publisher>
  <pubdate>2014</pubdate>
  <volume>13</volume>
  <issue>1</issue>
  <fpage>57</fpage>
  <lpage>-65</lpage>
</bibl>

<bibl id="B4">
  <title><p>{UK} {Biobank}: {An} {Open} {Access} {Resource} for {Identifying}
  the {Causes} of a {Wide} {Range} of {Complex} {Diseases} of {Middle} and
  {Old} {Age}</p></title>
  <aug>
    <au><snm>Sudlow</snm><fnm>C</fnm></au>
    <au><snm>Gallacher</snm><fnm>J</fnm></au>
    <au><snm>Allen</snm><fnm>N</fnm></au>
    <au><snm>Beral</snm><fnm>V</fnm></au>
    <au><snm>Burton</snm><fnm>P</fnm></au>
    <au><snm>Danesh</snm><fnm>J</fnm></au>
    <au><snm>Downey</snm><fnm>P</fnm></au>
    <au><snm>Elliott</snm><fnm>P</fnm></au>
    <au><snm>Green</snm><fnm>J</fnm></au>
    <au><snm>Landray</snm><fnm>M</fnm></au>
    <au><snm>Liu</snm><fnm>B</fnm></au>
    <au><snm>Matthews</snm><fnm>P</fnm></au>
    <au><snm>Ong</snm><fnm>G</fnm></au>
    <au><snm>Pell</snm><fnm>J</fnm></au>
    <au><snm>Silman</snm><fnm>A</fnm></au>
    <au><snm>Young</snm><fnm>A</fnm></au>
    <au><snm>Sprosen</snm><fnm>T</fnm></au>
    <au><snm>Peakman</snm><fnm>T</fnm></au>
    <au><snm>Collins</snm><fnm>R</fnm></au>
  </aug>
  <source>PLOS Medicine</source>
  <pubdate>2015</pubdate>
  <volume>12</volume>
  <issue>3</issue>
  <fpage>e1001779</fpage>
  <url>https://journals.plos.org/plosmedicine/article?id=10.1371/journal.pmed.1001779</url>
  <note>Publisher: Public Library of Science</note>
</bibl>

<bibl id="B5">
  <title><p>Multi-{Drug} {Featurization} and {Deep} {Learning} {Improve}
  {Patient}-{Specific} {Predictions} of {Adverse} {Events}</p></title>
  <aug>
    <au><snm>Anastopoulos</snm><fnm>IN</fnm></au>
    <au><snm>Herczeg</snm><fnm>CK</fnm></au>
    <au><snm>Davis</snm><fnm>KN</fnm></au>
    <au><snm>Dixit</snm><fnm>AC</fnm></au>
  </aug>
  <source>International Journal of Environmental Research and Public
  Health</source>
  <pubdate>2021</pubdate>
  <volume>18</volume>
  <issue>5</issue>
  <fpage>2600</fpage>
  <url>https://www.mdpi.com/1660-4601/18/5/2600</url>
  <note>Number: 5 Publisher: Multidisciplinary Digital Publishing
  Institute</note>
</bibl>

<bibl id="B6">
  <title><p>Patterns of {Multimorbidity} in {Middle}-{Aged} and {Older}
  {Adults}: {An} {Analysis} of the {UK} {Biobank} {Data}</p></title>
  <aug>
    <au><snm>Zemedikun</snm><fnm>DT</fnm></au>
    <au><snm>Gray</snm><fnm>LJ</fnm></au>
    <au><snm>Khunti</snm><fnm>K</fnm></au>
    <au><snm>Davies</snm><fnm>MJ</fnm></au>
    <au><snm>Dhalwani</snm><fnm>NN</fnm></au>
  </aug>
  <source>Mayo Clinic Proceedings</source>
  <pubdate>2018</pubdate>
  <volume>93</volume>
  <issue>7</issue>
  <fpage>857</fpage>
  <lpage>-866</lpage>
  <url>https://www.sciencedirect.com/science/article/pii/S0025619618301198</url>
</bibl>

<bibl id="B7">
  <title><p>Multimorbidity, polypharmacy, and {COVID}-19 infection within the
  {UK} {Biobank} cohort</p></title>
  <aug>
    <au><snm>McQueenie</snm><fnm>R</fnm></au>
    <au><snm>Foster</snm><fnm>HME</fnm></au>
    <au><snm>Jani</snm><fnm>BD</fnm></au>
    <au><snm>Katikireddi</snm><fnm>SV</fnm></au>
    <au><snm>Sattar</snm><fnm>N</fnm></au>
    <au><snm>Pell</snm><fnm>JP</fnm></au>
    <au><snm>Ho</snm><fnm>FK</fnm></au>
    <au><snm>Niedzwiedz</snm><fnm>CL</fnm></au>
    <au><snm>Hastie</snm><fnm>CE</fnm></au>
    <au><snm>Anderson</snm><fnm>J</fnm></au>
    <au><snm>Mark</snm><fnm>PB</fnm></au>
    <au><snm>Sullivan</snm><fnm>M</fnm></au>
    <au><snm>O’Donnell</snm><fnm>CA</fnm></au>
    <au><snm>Mair</snm><fnm>FS</fnm></au>
    <au><snm>Nicholl</snm><fnm>BI</fnm></au>
  </aug>
  <source>PLOS ONE</source>
  <pubdate>2020</pubdate>
  <volume>15</volume>
  <issue>8</issue>
  <fpage>e0238091</fpage>
  <url>https://journals.plos.org/plosone/article?id=10.1371/journal.pone.0238091</url>
  <note>Publisher: Public Library of Science</note>
</bibl>

<bibl id="B8">
  <title><p>Genome-wide association study of medication-use and associated
  disease in the {UK} {Biobank}</p></title>
  <aug>
    <au><snm>Wu</snm><fnm>Y</fnm></au>
    <au><snm>Byrne</snm><fnm>EM</fnm></au>
    <au><snm>Zheng</snm><fnm>Z</fnm></au>
    <au><snm>Kemper</snm><fnm>KE</fnm></au>
    <au><snm>Yengo</snm><fnm>L</fnm></au>
    <au><snm>Mallett</snm><fnm>AJ</fnm></au>
    <au><snm>Yang</snm><fnm>J</fnm></au>
    <au><snm>Visscher</snm><fnm>PM</fnm></au>
    <au><snm>Wray</snm><fnm>NR</fnm></au>
  </aug>
  <source>Nature Communications</source>
  <pubdate>2019</pubdate>
  <volume>10</volume>
  <issue>1</issue>
  <fpage>1891</fpage>
  <url>https://www.nature.com/articles/s41467-019-09572-5</url>
  <note>Bandiera\_abtest: a Cc\_license\_type: cc\_by Cg\_type: Nature Research
  Journals Number: 1 Primary\_atype: Research Publisher: Nature Publishing
  Group Subject\_term: Genetics research;Genome-wide association
  studies;Genotype Subject\_term\_id:
  genetics-research;genome-wide-association-studies;genotype</note>
</bibl>

<bibl id="B9">
  <title><p>Pharmacogenetics at {Scale}: {An} {Analysis} of the {UK}
  {Biobank}</p></title>
  <aug>
    <au><snm>McInnes</snm><fnm>G</fnm></au>
    <au><snm>Lavertu</snm><fnm>A</fnm></au>
    <au><snm>Sangkuhl</snm><fnm>K</fnm></au>
    <au><snm>Klein</snm><fnm>TE</fnm></au>
    <au><snm>Whirl Carrillo</snm><fnm>M</fnm></au>
    <au><snm>Altman</snm><fnm>RB</fnm></au>
  </aug>
  <source>Clinical Pharmacology \& Therapeutics</source>
  <pubdate>2021</pubdate>
  <volume>109</volume>
  <issue>6</issue>
  <fpage>1528</fpage>
  <lpage>-1537</lpage>
  <url>https://onlinelibrary.wiley.com/doi/abs/10.1002/cpt.2122</url>
  <note>\_eprint:
  https://onlinelibrary.wiley.com/doi/pdf/10.1002/cpt.2122</note>
</bibl>

<bibl id="B10">
  <title><p>Therapeutics Data Commons: Machine Learning Datasets and Tasks for
  Drug Discovery and Development</p></title>
  <aug>
    <au><snm>Huang</snm><fnm>K</fnm></au>
    <au><snm>Fu</snm><fnm>T</fnm></au>
    <au><snm>Gao</snm><fnm>W</fnm></au>
    <au><snm>Zhao</snm><fnm>Y</fnm></au>
    <au><snm>Roohani</snm><fnm>Y</fnm></au>
    <au><snm>Leskovec</snm><fnm>J</fnm></au>
    <au><snm>Coley</snm><fnm>CW</fnm></au>
    <au><snm>Xiao</snm><fnm>C</fnm></au>
    <au><snm>Sun</snm><fnm>J</fnm></au>
    <au><snm>Zitnik</snm><fnm>M</fnm></au>
  </aug>
  <source>Proceedings of Neural Information Processing Systems, NeurIPS
  Datasets and Benchmarks</source>
  <pubdate>2021</pubdate>
</bibl>

<bibl id="B11">
  <title><p>The incidence of thyroid disorders in the community: a twenty-year
  follow-up of the Whickham Survey</p></title>
  <aug>
    <au><snm>Vanderpump</snm><fnm>MPJ</fnm></au>
    <au><snm>Tunbrldge</snm><fnm>WMG</fnm></au>
    <au><snm>French</snm><fnm>Ja</fnm></au>
    <au><snm>Appleton</snm><fnm>D</fnm></au>
    <au><snm>Bates</snm><fnm>D</fnm></au>
    <au><snm>Clark</snm><fnm>F</fnm></au>
    <au><snm>Evans</snm><fnm>JG</fnm></au>
    <au><snm>Hasan</snm><fnm>DM</fnm></au>
    <au><snm>Rodgers</snm><fnm>H</fnm></au>
    <au><snm>Tunbridge</snm><fnm>F</fnm></au>
    <au><cnm>others</cnm></au>
  </aug>
  <source>Clinical endocrinology</source>
  <publisher>Wiley Online Library</publisher>
  <pubdate>1995</pubdate>
  <volume>43</volume>
  <issue>1</issue>
  <fpage>55</fpage>
  <lpage>-68</lpage>
</bibl>

<bibl id="B12">
  <title><p>Beta-blocker efficacy across different cardiovascular indications:
  an umbrella review and meta-analytic assessment</p></title>
  <aug>
    <au><snm>Ziff</snm><fnm>OJ</fnm></au>
    <au><snm>Samra</snm><fnm>M</fnm></au>
    <au><snm>Howard</snm><fnm>JP</fnm></au>
    <au><snm>Bromage</snm><fnm>DI</fnm></au>
    <au><snm>Ruschitzka</snm><fnm>F</fnm></au>
    <au><snm>Francis</snm><fnm>DP</fnm></au>
    <au><snm>Kotecha</snm><fnm>D</fnm></au>
  </aug>
  <source>BMC medicine</source>
  <publisher>Springer</publisher>
  <pubdate>2020</pubdate>
  <volume>18</volume>
  <issue>1</issue>
  <fpage>1</fpage>
  <lpage>-11</lpage>
</bibl>

<bibl id="B13">
  <title><p>Efficacy and safety of statin therapy in older people: a
  meta-analysis of individual participant data from 28 randomised controlled
  trials</p></title>
  <aug>
    <au><snm>Armitage</snm><fnm>J</fnm></au>
    <au><snm>Baigent</snm><fnm>C</fnm></au>
    <au><snm>Barnes</snm><fnm>E</fnm></au>
    <au><snm>Betteridge</snm><fnm>DJ</fnm></au>
    <au><snm>Blackwell</snm><fnm>L</fnm></au>
    <au><snm>Blazing</snm><fnm>M</fnm></au>
    <au><snm>Bowman</snm><fnm>L</fnm></au>
    <au><snm>Braunwald</snm><fnm>E</fnm></au>
    <au><snm>Byington</snm><fnm>R</fnm></au>
    <au><snm>Cannon</snm><fnm>C</fnm></au>
    <au><cnm>others</cnm></au>
  </aug>
  <source>The Lancet</source>
  <publisher>Elsevier</publisher>
  <pubdate>2019</pubdate>
  <volume>393</volume>
  <issue>10170</issue>
  <fpage>407</fpage>
  <lpage>-415</lpage>
</bibl>

<bibl id="B14">
  <title><p>{CPIC}® {Guideline} for {Simvastatin} and {SLCO1B1}</p></title>
  <url>https://cpicpgx.org/guidelines/guideline-for-simvastatin-and-slco1b1/</url>
</bibl>

<bibl id="B15">
  <title><p>statsmodels: Econometric and statistical modeling with
  python</p></title>
  <aug>
    <au><snm>Seabold</snm><fnm>S</fnm></au>
    <au><snm>Perktold</snm><fnm>J</fnm></au>
  </aug>
  <source>9th Python in Science Conference</source>
  <pubdate>2010</pubdate>
</bibl>

<bibl id="B16">
  <title><p>RAPIDS: Collection of Libraries for End to End GPU Data
  Science</p></title>
  <aug>
    <au><snm>Team</snm><fnm>RD</fnm></au>
  </aug>
  <pubdate>2018</pubdate>
  <url>https://rapids.ai</url>
</bibl>

<bibl id="B17">
  <title><p>ukbREST: efficient and streamlined data access for reproducible
  research in large biobanks</p></title>
  <aug>
    <au><snm>Pividori</snm><fnm>M</fnm></au>
    <au><snm>Im</snm><fnm>HK</fnm></au>
  </aug>
  <source>Bioinformatics</source>
  <publisher>Oxford University Press</publisher>
  <pubdate>2019</pubdate>
  <volume>35</volume>
  <issue>11</issue>
  <fpage>1971</fpage>
  <lpage>-1973</lpage>
</bibl>

<bibl id="B18">
  <title><p>Statistical methods for research workers</p></title>
  <aug>
    <au><snm>Fisher</snm><fnm>RA</fnm></au>
  </aug>
  <source>Edinburgh: Oliver and Boyd, 1934 and The logic of inductive
  interence; Royal Statistical Society</source>
  <pubdate>1935</pubdate>
  <volume>98</volume>
  <fpage>S</fpage>
  <lpage>-39</lpage>
</bibl>

<bibl id="B19">
  <title><p>Basic Statistics and Data Analysis. 2003</p></title>
  <aug>
    <au><snm>Kitchens</snm><fnm>LJ</fnm></au>
  </aug>
  <source>Pacific Grove, CA: Thomson Brooks/Cole</source>
  <pubdate>2002</pubdate>
</bibl>

<bibl id="B20">
  <title><p>Mathematical methods of organizing and planning
  production</p></title>
  <aug>
    <au><snm>Kantorovich</snm><fnm>LV</fnm></au>
  </aug>
  <source>Management science</source>
  <publisher>INFORMS</publisher>
  <pubdate>1960</pubdate>
  <volume>6</volume>
  <issue>4</issue>
  <fpage>366</fpage>
  <lpage>-422</lpage>
</bibl>

</refgrp>
} 

\clearpage

\section*{Supplementary Methods}

\subsection*{Statistical tests}

\subsubsection*{Fisher's exact test}

As to find significantly enriched Meta-Drugs or medications in clusters we performed one-sided Fisher's exact test \cite{fisher1935statistical} according to the values in Table \ref{table:contingency-table}.

$H_0:$ The input table is from the hypergeometric distribution with parameters:\\ 
\phantom{}$\qquad \qquad \qquad \qquad \qquad \qquad \qquad M = a + b + c + d$, $n = a + b$ and $N = a + c$ \\
$H_a:$ A random table has $x >= a$

\bgroup
\def\arraystretch{1.5}
\begin{table}[ht]
\begin{center}
    \normalsize
\begin{tabular}{l|l|c|c|c}
\cline{3-4}
\multicolumn{2}{c|}{}&$\#P$ in cluster&$\#P$ not in cluster&\multicolumn{1}{c}{}\\
\cline{2-4}
\multirow{2}{*} \phantom{} & $\#P$ taking medication & $a$ & $b$ & \phantom{}\\
\cline{2-4} \phantom{} & $\#P$ not taking medication & $c$ & $d$ & \\
\cline{2-4}
\end{tabular}
    \end{center}
    \caption{Contingency table.\label{table:contingency-table}}
\end{table}
\egroup

where, \\
$\#P$: Number of Participants\\

\subsubsection*{Two-Proportions Z-Test}

We utilize the two-proportions Z-test \cite{kitchens2002basic} in order to examine whether the proportions of an attribute of two groups (clusters) are statistically different or not. For this we calculate the test statistic, $Z$, according to equation \ref{eq:z-statistic}.

$H_0: p_A = p_B$ \\
$H_a: p_A \neq p_B$

\begin{equation}
 Z = \frac{p_A - p_B}{\sqrt{p(1-p)(\frac{1}{n_A}+\frac{1}{n_B})}}
\label{eq:z-statistic}
\end{equation}

where, \\
$p_A$ is the proportion observed in group $A$ with size $n_A$, \\
$p_B$ is the proportion observed in group $B$ with size $n_B$, \\
$p$ is the overall proportion, \\
$n_A$ is the total number of elements in $A$, \\
$n_B$ is the total number of elements in $B$.

\subsubsection*{Two-Sample Kolmogorov-Smirnov Test (Continuous distributions)}

In order to compare two continuous distributions we utilize the two-sample Kolmogorov-Smirnov test, where we calculate the test statistic, $D$, according to equation \ref{eq:KS-statistic}.

$H_0: F_{n_A} = F_{n_B}$ \\
$H_a: F_{n_A} \neq F_{n_B}$

\begin{equation}
 D = Maximum|F_{n_A}(X)-F_{n_B}(X)|
\label{eq:KS-statistic}
\end{equation}

where, \\
$n_A$ are the observations from the first sample (cluster), \\
$n_B$ are the observations from the second sample (cluster), \\ 
$F_{n_A}$ is the cumulative frequency distribution of a random sample of $n_A$, \\
$F_{n_B}$ is the cumulative frequency distribution of a random sample of $n_B$, \\

\subsubsection*{P-value correction for multiple tests}

After p-value calculations for each cluster attribute according to Table \ref{table:stattests} we performed a Bonferroni p-value correction ($\alpha = 0.05$) for each of the attributes separately.

Additionally, after p-value calculations (Fisher's exact test) for Meta-Drug or medications enrichment between clusters the same Bonferroni correction was performed.

\subsection*{Distance calculations}

\subsubsection*{Wasserstein distance (Discrete distributions)}

We utilized the Wasserstein distance \cite{kantorovich1960mathematical}, $W$ (equation \ref{eq:Wasserstein}), where a statistical test was not possible, in order to have a measure of closeness of two discrete distributions. This distance represents the minimum amount of "work" required to transform the distribution of values $n_A$ into the distribution of values $n_B$.

\begin{equation}
 W = \sum_{i=1}^{k}{|F_{n_A}-F_{n_B}|\times\delta_i}
\label{eq:Wasserstein}
\end{equation}

where, \\
$\delta$ is a vector of differences between pairs of successive values of $n_A$ and $n_B$. \\

Instead of a statistical significance we set a threshold at the first quartile, $Q1$ ($25$th percentile), of all the Wasserstein distances above which we consider two discrete distributions to be dissimilar.

\clearpage
\section*{Supplementary Files}

\beginsupplement

\subsection*{Supplementary Tables}

\begin{table}[ht]
\begin{center}
    \normalsize
\begin{tabular}{l|r|r|r|r|r}
\textbf{Cluster} & \textbf{\# Participants} & \textbf{\% Female} & \textbf{Avg. age} & \textbf{Avg. \# drugs} & \textbf{\% Five-year mortality}\\
\midrule
     C0 &           $1'906$ &  $43.2$ & $60.7$ &       $1.9$ &         2.1 \\
     C1 &           $4'202$ &  $68.6$ & $55.2$ &       $1.7$ &         1.3 \\
     C2 &          $21'561$ &  $29.6$ & $61.6$ &       $4.3$ &         2.3 \\
     C3 &          $30'500$ &  $64.8$ & $52.9$ &       $1.1$ &         0.8 \\
     C4 &          $10'865$ &  $59.3$ & $59.6$ &       $2.2$ &         1.6 \\
     C5 &          $11'588$ &  $52.6$ & $54.4$ &       $1.8$ &         1.1 \\
     C6 &           $8'316$ &  $62.3$ & $57.6$ &       $1.8$ &         0.7 \\
     C7 &          $28'361$ &  $67.0$ & $56.2$ &       $2.9$ &         1.5 \\
     C8 &          $14'652$ &  $49.3$ & $60.4$ &       $2.9$ &         1.6 \\
     C9 &          $21'972$ &  $55.7$ & $56.4$ &       $3.1$ &         2.8 \\
    C10 &          $10'298$ &  $88.7$ & $60.3$ &       $3.8$ &         2.4 \\
    C11 &            $786$ &  $69.1$ & $57.1$ &       $2.1$ &         2.6 \\
    C12 &           $6'437$ &  $41.0$ & $58.1$ &       $1.0$ &         1.5 \\
    C13 &           $2'342$ &  $84.0$ & $54.1$ &       $1.6$ &         0.6 \\
    C14 &          $19'760$ &  $62.2$ & $57.2$ &       $4.5$ &         2.4 \\
    C15 &           $4'387$ &  $60.7$ & $56.6$ &       $1.9$ &         1.7 \\
    C16 &          $10'465$ &  $63.6$ & $59.1$ &       $3.2$ &         1.4 \\
    C17 &           $7'267$ &  $40.9$ & $58.2$ &       $1.5$ &         1.6 \\
    C18 &           $7'576$ &  $73.3$ & $52.9$ &       $1.5$ &         0.9 \\
    C19 &          $23'924$ &  $36.2$ & $61.4$ &       $3.6$ &         2.5 \\
    C20 &           $7'011$ &  $85.8$ & $56.1$ &       $1.1$ &         0.9 \\
    C21 &           $3'665$ &  $57.6$ & $53.7$ &       $1.4$ &         0.8 \\
    C22 &          $31'124$ &  $48.7$ & $61.1$ &       $6.3$ &         4.1 \\
    C23 &           $1'596$ &  $47.3$ & $55.2$ &       $1.2$ &         1.1 \\
    C24 &           $7'470$ &  $58.7$ & $57.8$ &       $1.0$ &         0.8 \\
    C25 &           $4'924$ &  $91.6$ & $59.8$ &       $1.5$ &         1.0 \\
    C26 &           $1'034$ &  $98.5$ & $56.1$ &       $1.2$ &         5.2 \\
    C27 &           $2'972$ &  $26.4$ & $58.6$ &       $1.4$ &         1.2 \\
    C28 &           $1'139$ &  $53.6$ & $53.8$ &       $1.3$ &         1.2 \\
    C29 &           $8'612$ &  $49.5$ & $57.2$ &       $1.3$ &         1.7 \\
    C30 &           $5'650$ &  $37.8$ & $59.7$ &       $1.0$ &         1.1 \\
    Excluded &      $5'823$ &   -     &   -    &   -         &  -          \\
\end{tabular}
    \end{center}
    \caption{Cluster characteristics for the main clusters. Some participants were excluded from the cluster analysis due to missing values.\label{table:c31_summary_stats}}
\end{table}

\begin{sidewaystable}
\centering
\begin{tabularx}{\textheight}{l|X|X|X|X|X}
\textbf{Cluster} & \textbf{Rank $1$ medication} & \textbf{Rank $2$ medication} & \textbf{Rank $3$ medication} & \textbf{Rank $4$ medication} & \textbf{Rank $5$ medication}
\\\hline
C0 & Latanoprost & Timolol & Bimatoprost & Timolol / Dorzolamide & Travoprost \\
C1 & Amitriptyline & Venlafaxine & Dosulepin & Mirtazapine & St. John'S Wort \\
C2 & Acetylsalicylic Acid & Simvastatin & Ramipril & Bendroflumethiazide & Amlodipine \\
C3 & Acetaminophen & Glucosamine & Tramadol & Chondroitin Sulfate & Naproxen \\
C4 & Bendroflumethiazide & Atenolol & Amlodipine & Ramipril & Doxazosin \\
C5 & Salbutamol & Beclomethasone Dipropionate & Salmeterol / Fluticasone Propionate & Fluticasone Propionate & Formoterol / Budesonide \\
C6 & Fish Oil & Glucosamine & Chondroitin Sulfate & Acetaminophen & Ezetimibe \\
C7 & Acetaminophen & Glucosamine & Acetylsalicylic Acid & Omeprazole & Levothyroxine \\
C8 & Simvastatin & Acetaminophen & Atorvastatin & Levothyroxine & Omeprazole \\
C9 & Omeprazole & Acetaminophen & Levothyroxine & Folic Acid & Acetaminophen / Codeine \\
C10 & Cholecalciferol / Calcium Carbonate & Alendronic Acid & Acetaminophen & Glucosamine & Levothyroxine \\
C11 & Zopiclone & Acetaminophen & Temazepam & Zolpidem & Citalopram \\
C12 & Acetylsalicylic Acid & Verapamil & Diltiazem & Clopidogrel & Nitroglycerin \\
C13 & Sumatriptan & Acetaminophen & Zolmitriptan & Naratriptan & Rizatriptan \\
C14 & Salbutamol & Beclomethasone Dipropionate & Acetaminophen & Salmeterol / Fluticasone Propionate & Acetylsalicylic Acid \\
C15 & Acetaminophen / Codeine & Acetaminophen & Acetaminophen / Dihydrocodeine & Omeprazole & Amitriptyline \\
C16 & Acetaminophen & Bendroflumethiazide & Amlodipine & Glucosamine & Atenolol \\
C17 & Amlodipine & Ramipril & Lisinopril & Perindopril & Felodipine \\
C18 & Citalopram & Fluoxetine & Acetaminophen & Paroxetine & Sertraline \\
C19 & Acetylsalicylic Acid & Simvastatin & Atorvastatin & Metformin & Atenolol \\
C20 & Levothyroxine & Iron & Cyanocobalamin & Hydroxocobalamin & Insulin \\
C21 & Cetirizine & Loratadine & Fexofenadine & Chlorphenamine & Desloratadine \\
C22 & Acetylsalicylic Acid & Simvastatin & Acetaminophen & Bendroflumethiazide & Omeprazole \\
C23 & Ranitidine & Cimetidine & Sodium Bicarbonate / Calcium Carbonate / Alginic Acid & Levothyroxine & Allopurinol \\
C24 & Glucosamine & Chondroitin Sulfate & Naproxen & Diclofenac / Misoprostol & Meloxicam \\
C25 & Cholecalciferol / Calcium Carbonate & Alendronic Acid & Calcium & Calcium Carbonate & Tibolone \\
C26 & Tamoxifen & Anastrozole & Letrozole & Exemestane & Levothyroxine \\
C27 & Tamsulosin & Iron & Finasteride & Alfuzosin & Dutasteride \\
C28 & Oxytetracycline & Doxycycline & Lymecycline & Metronidazole & Minocycline \\
C29 & Omeprazole & Lansoprazole & Sodium Bicarbonate / Calcium Carbonate / Alginic Acid & Esomeprazole & Levothyroxine \\
C30 & Simvastatin & Atorvastatin & Rosuvastatin & Pravastatin & Fenofibrate \\
\end{tabularx}
    \caption{Top 5 medications per main cluster.\label{table:c31_top5_meds}}
\end{sidewaystable}

\begin{sidewaystable}
\centering
\begin{tabularx}{\textheight}{l|X|X|X|X|X}
\textbf{Cluster} & \textbf{Rank $1$ disease} & \textbf{Rank $2$ disease} & \textbf{Rank $3$ disease} & \textbf{Rank $4$ disease} & \textbf{Rank $5$ disease}
\\\hline
C0 & glaucoma & hypertension & cataract & eye/eyelid problem & osteoarthritis \\
C1 & depression & hypertension & anxiety/panic attacks & osteoarthritis & asthma \\
C2 & hypertension & high cholesterol & heart attack/myocardial infarction & angina & diabetes \\
C3 & hypertension & osteoarthritis & asthma & hayfever/allergic rhinitis & migraine \\
C4 & hypertension & osteoarthritis & asthma & high cholesterol & hypothyroidism/myxoedema \\
C5 & asthma & hayfever/allergic rhinitis & hypertension & eczema/dermatitis & emphysema/chronic bronchitis \\
C6 & hypertension & osteoarthritis & high cholesterol & asthma & hayfever/allergic rhinitis \\
C7 & osteoarthritis & hypertension & hypothyroidism/myxoedema & gastro-oesophageal reflux (gord) / gastric reflux & depression \\
C8 & high cholesterol & hypertension & diabetes & osteoarthritis & hypothyroidism/myxoedema \\
C9 & hypertension & depression & osteoarthritis & asthma & hypothyroidism/myxoedema \\
C10 & osteoporosis & hypertension & osteoarthritis & hypothyroidism/myxoedema & breast cancer \\
C11 & depression & hypertension & insomnia & asthma & osteoarthritis \\
C12 & hypertension & asthma & osteoarthritis & hayfever/allergic rhinitis & high cholesterol \\
C13 & migraine & hypertension & asthma & osteoarthritis & hayfever/allergic rhinitis \\
C14 & asthma & hypertension & hayfever/allergic rhinitis & osteoarthritis & high cholesterol \\
C15 & osteoarthritis & hypertension & depression & back problem & asthma \\
C16 & hypertension & osteoarthritis & high cholesterol & asthma & migraine \\
C17 & hypertension & asthma & osteoarthritis & hayfever/allergic rhinitis & high cholesterol \\
C18 & depression & anxiety/panic attacks & hypertension & asthma & osteoarthritis \\
C19 & high cholesterol & hypertension & angina & diabetes & heart attack/myocardial infarction \\
C20 & hypothyroidism/myxoedema & hyperthyroidism/thyrotoxicosis & hypertension & asthma & osteoarthritis \\
C21 & hayfever/allergic rhinitis & asthma & hypertension & eczema/dermatitis & allergy/hypersensitivity/anaphylaxis \\
C22 & hypertension & high cholesterol & diabetes & osteoarthritis & angina \\
C23 & gastro-oesophageal reflux (gord) / gastric reflux & hypertension & hiatus hernia & asthma & osteoarthritis \\
C24 & osteoarthritis & hypertension & asthma & hayfever/allergic rhinitis & eczema/dermatitis \\
C25 & osteoporosis & hypertension & osteopenia & osteoarthritis & breast cancer \\
C26 & breast cancer & hypertension & asthma & osteoarthritis \\
C27 & enlarged prostate & iron deficiency anaemia & hypertension & bph / benign prostatic hypertrophy & asthma \\
C28 & rosacea & hypertension & asthma & eczema/dermatitis & acne/acne vulgaris \\
C29 & gastro-oesophageal reflux (gord) / gastric reflux & hiatus hernia & hypertension & osteoarthritis & asthma \\
C30 & high cholesterol & hypertension & osteoarthritis & asthma & diabetes \\
\end{tabularx}
    \caption{Top 5 medical conditions per main cluster.\label{table:c31_top5_medconds}}
\end{sidewaystable}

\begin{sidewaystable}
\centering
\begin{tabularx}{\textheight}{l|X|X|X|X|X}
\textbf{Cluster} & \textbf{Rank $1$ Meta-Drug} & \textbf{Rank $2$ Meta-Drug} & \textbf{Rank $3$ Meta-Drug} & \textbf{Rank $4$ Meta-Drug} & \textbf{Rank $5$ Meta-Drug}
\\\hline
C0 & glaucoma & hypertension & osteoarthritis & angina & high cholesterol \\
C1 & depression & osteoarthritis & gord / gastric reflux & hypertension & hypothyroidism/myxoedema \\
C2 & hypertension & high cholesterol & angina & diabetes & osteoarthritis \\
C3 & osteoarthritis & - & - & - & - \\
C4 & hypertension & angina & hypothyroidism/myxoedema & gord / gastric reflux & high cholesterol \\
C5 & asthma & hayfever/allergic rhinitis & gord / gastric reflux & hypertension & depression \\
C6 & high cholesterol & osteoarthritis & hypertension & gord / gastric reflux & osteoporosis \\
C7 & osteoarthritis & gord / gastric reflux & angina & depression & hypothyroidism/myxoedema \\
C8 & high cholesterol & hypertension & osteoarthritis & gord / gastric reflux & diabetes \\
C9 & gord / gastric reflux & depression & osteoarthritis & rheumatoid arthritis & irritable bowel syndrome \\
C10 & osteoporosis & osteoarthritis & high cholesterol & gord / gastric reflux & hypertension \\
C11 & insomnia & depression & osteoarthritis & gord / gastric reflux & - \\
C12 & angina & - & - & - & - \\
C13 & migraine & osteoarthritis & hypertension & depression & gord / gastric reflux \\
C14 & asthma & osteoarthritis & hypertension & gord / gastric reflux & high cholesterol \\
C15 & back problem & osteoarthritis & depression & gord / gastric reflux & hypothyroidism/myxoedema \\
C16 & hypertension & osteoarthritis & high cholesterol & angina & depression \\
C17 & hypertension & gord / gastric reflux & gout & asthma & hypothyroidism/myxoedema \\
C18 & depression & osteoarthritis & gord / gastric reflux & hypertension & hypothyroidism/myxoedema \\
C19 & angina & high cholesterol & hypertension & osteoarthritis & diabetes \\
C20 & hypothyroidism/myxoedema & iron deficiency anaemia & pernicious anaemia & diabetes & irritable bowel syndrome \\
C21 & hayfever/allergic rhinitis & asthma & osteoarthritis & hypothyroidism/myxoedema & gord / gastric reflux \\
C22 & hypertension & high cholesterol & angina & gord / gastric reflux & diabetes \\
C23 & gord / gastric reflux & hypothyroidism/myxoedema & irritable bowel syndrome & gout & eczema/dermatitis \\
C24 & osteoarthritis  & - & - & - & - \\
C25 & osteoporosis & high cholesterol & hypothyroidism/myxoedema & gord / gastric reflux & - \\
C26 & breast cancer & osteoarthritis & hypothyroidism/myxoedema & hypertension & gord / gastric reflux \\
C27 & enlarged prostate & iron deficiency anaemia & osteoarthritis & gord / gastric reflux & hypertension \\
C28 & rosacea & acne/acne vulgaris & hypertension & gord / gastric reflux & osteoarthritis \\
C29 & gord / gastric reflux & hypothyroidism/myxoedema & irritable bowel syndrome & hayfever/allergic rhinitis & gout \\
C30 & high cholesterol  & - & - & - & - \\
\end{tabularx}
    \caption{Top 5 Meta-Drugs per main cluster.\label{table:c31_top5_metadrugs}}
\end{sidewaystable}

\begin{table}
\centering
\begin{tabularx}{\textheight}{l|l| l| p{0.45\linewidth}}
\textbf{Cluster} &                \textbf{Top $1$-combination} &                                   \textbf{Top $2$-combination} &                                              \textbf{Top $3$-combination} \\ \hline
     C0 &          latanoprost &                    latanoprost, timolol &                    dorzolamide, latanoprost, timolol \\
     C1 &        amitriptyline &            acetaminophen, amitriptyline &                acetaminophen, amitriptyline, codeine \\
     C2 & acetylsalicylic acid &       acetylsalicylic acid, simvastatin &          acetylsalicylic acid, ramipril, simvastatin \\
     C3 &        acetaminophen &              acetaminophen, glucosamine &               acetaminophen, diclofenac, misoprostol \\
     C4 &  bendroflumethiazide &           atenolol, bendroflumethiazide &            amlodipine, atenolol, bendroflumethiazide \\
     C5 &           salbutamol & beclomethasone dipropionate, salbutamol &       fluticasone propionate, salbutamol, salmeterol \\
     C6 &          glucosamine &      calcium carbonate, cholecalciferol &                  acetaminophen, codeine, glucosamine \\
     C7 &        acetaminophen &     acetaminophen, acetylsalicylic acid &     acetaminophen, acetylsalicylic acid, glucosamine \\
     C8 &          simvastatin &              acetaminophen, simvastatin &                  acetaminophen, codeine, simvastatin \\
     C9 &        acetaminophen &                  acetaminophen, codeine &              acetaminophen, folic acid, methotrexate \\
    C10 &    calcium carbonate &      calcium carbonate, cholecalciferol &    acetaminophen, calcium carbonate, cholecalciferol \\
    C11 &            zopiclone &                acetaminophen, zopiclone &                    acetaminophen, codeine, zopiclone \\
    C12 & acetylsalicylic acid &         acetylsalicylic acid, diltiazem &          acetylsalicylic acid, diltiazem, nicorandil \\
    C13 &          sumatriptan &                  acetaminophen, codeine &                    acetaminophen, buclizine, codeine \\
    C14 &           salbutamol &               acetaminophen, salbutamol &       fluticasone propionate, salbutamol, salmeterol \\
    C15 &        acetaminophen &                  acetaminophen, codeine &                   acetaminophen, codeine, omeprazole \\
    C16 &        acetaminophen &      acetaminophen, bendroflumethiazide &       acetaminophen, amlodipine, bendroflumethiazide \\
    C17 &           amlodipine &                    amlodipine, ramipril &            amlodipine, bendroflumethiazide, ramipril \\
    C18 &           citalopram &               acetaminophen, citalopram &       calcium carbonate, cholecalciferol, citalopram \\
    C19 & acetylsalicylic acid &       acetylsalicylic acid, simvastatin &     acetaminophen, acetylsalicylic acid, simvastatin \\
    C20 &        levothyroxine &           cyanocobalamin, levothyroxine & calcium carbonate, levothyroxine, sodium bicarbonate \\
    C21 &           cetirizine & beclomethasone dipropionate, loratadine &       fluticasone propionate, loratadine, salmeterol \\
    C22 & acetylsalicylic acid &       acetylsalicylic acid, simvastatin &     acetaminophen, acetylsalicylic acid, simvastatin \\
    C23 &           ranitidine &        alginic acid, sodium bicarbonate &  alginic acid, calcium carbonate, sodium bicarbonate \\
    C24 &          glucosamine &                   glucosamine, naproxen &                 diclofenac, glucosamine, misoprostol \\
    C25 &    calcium carbonate &      calcium carbonate, cholecalciferol &  alendronic acid, calcium carbonate, cholecalciferol \\
    C26 &            tamoxifen &                levothyroxine, tamoxifen &               cinchocaine, hydrocortisone, tamoxifen \\
    C27 &           tamsulosin &                 finasteride, tamsulosin &               acetaminophen, finasteride, tamsulosin \\
    C28 &      oxytetracycline &          levothyroxine, oxytetracycline &  calcium carbonate, cholecalciferol, oxytetracycline \\
    C29 &           omeprazole &        alginic acid, sodium bicarbonate &  alginic acid, calcium carbonate, sodium bicarbonate \\
    C30 &          simvastatin &                fenofibrate, simvastatin &                                                 None \\
\end{tabularx}
    \caption{Top 3 drug combinations per main cluster.\label{table:c31_top3_drugcombs}}
\end{table}

\begin{table}
\centering
\begin{tabularx}{\textheight}{l|l| l| p{0.45\linewidth}}
\textbf{Cluster} &                \textbf{Top $1$-combination} &                                   \textbf{Top $2$-combination} &                                              \textbf{Top $3$-combination} \\ \hline

     C0 &          glaucoma &                  glaucoma, hypertension &                  angina, glaucoma, high cholesterol \\
     C1 &        depression &              depression, osteoarthritis &                    depression, gord, osteoarthritis \\
     C2 &      hypertension &          high cholesterol, hypertension &              angina, high cholesterol, hypertension \\
     C3 &    osteoarthritis &                                    None &                                                None \\
     C4 &      hypertension &                    angina, hypertension &                angina, hypertension, hypothyroidism \\
     C5 &            asthma &                        asthma, hayfever &                                                None \\
     C6 &  high cholesterol &        high cholesterol, osteoarthritis &      high cholesterol, hypertension, osteoarthritis \\
     C7 &    osteoarthritis &                    gord, osteoarthritis &                    depression, gord, osteoarthritis \\
     C8 &  high cholesterol &          high cholesterol, hypertension &      high cholesterol, hypertension, osteoarthritis \\
     C9 &              gord &                        depression, gord &                    depression, gord, osteoarthritis \\
    C10 &      osteoporosis &            osteoarthritis, osteoporosis &      high cholesterol, osteoarthritis, osteoporosis \\
    C11 &          insomnia &                    depression, insomnia &                depression, insomnia, osteoarthritis \\
    C12 &            angina &                                    None &                                                None \\
    C13 &          migraine &                migraine, osteoarthritis &                depression, migraine, osteoarthritis \\
    C14 &            asthma &                  asthma, osteoarthritis &                        asthma, gord, osteoarthritis \\
    C15 &      back problem &            back problem, osteoarthritis &            back problem, depression, osteoarthritis \\
    C16 &      hypertension &            hypertension, osteoarthritis &      high cholesterol, hypertension, osteoarthritis \\
    C17 &      hypertension &                      gord, hypertension &                                                None \\
    C18 &        depression &              depression, osteoarthritis &                                                None \\
    C19 &            angina &                angina, high cholesterol &              angina, high cholesterol, hypertension \\
    C20 &    hypothyroidism & hypothyroidism, iron deficiency anaemia &                                                None \\
    C21 &          hayfever &                        asthma, hayfever &                    asthma, hayfever, osteoarthritis \\
    C22 &      hypertension &          high cholesterol, hypertension &              angina, high cholesterol, hypertension \\
    C23 &              gord &                    gord, hypothyroidism &                                                None \\
    C24 &    osteoarthritis &                                    None &                                                None \\
    C25 &      osteoporosis &          high cholesterol, osteoporosis &                                                None \\
    C26 &     breast cancer &           breast cancer, osteoarthritis &                  breast cancer, bronchiectasis, UTI \\
    C27 & enlarged prostate &       enlarged prostate, osteoarthritis & enlarged prostate, high cholesterol, osteoarthritis \\
    C28 &           rosacea &                   hypertension, rosacea &                                                None \\
    C29 &              gord &                    gord, hypothyroidism &                                                None \\
    C30 &  high cholesterol &                                    None &                                                                 None \\

\end{tabularx}
    \caption{Top 3 Meta-Drug combinations per main cluster.\label{table:c31_top3_metacombs}}
\end{table}

\begin{table}[ht]
\begin{center}
    \normalsize
\begin{tabular}{l|l}
\textbf{Original category} & \textbf{Mapping}
\\\hline
British &  European  \\
Any other white background &  European  \\
Irish &  European  \\
White &  European  \\
Indian &  Asian  \\
Pakistani &  Asian  \\
Any other Asian background &  Asian  \\
Chinese &  Asian  \\
Bangladeshi &  Asian  \\
Asian or Asian British &  Asian  \\
Caribbean &  Carribean  \\
African &  African  \\
Black or Black British &  African \\
Any other Black background &  African  \\
Any other mixed background &  Other  \\
White and Asian &  Other  \\
White and Black Caribbean &  Other  \\
White and Black African &  Other  \\
Mixed &  Other  \\
Other ethnic group &  Other  \\ 
Prefer not to answer &  NA  \\
Do not know &  NA  \\
\end{tabular}
    \end{center}
    \caption{Mapping of Ethnic background values.\label{table:ethnic_back_mapping}}
\end{table}

\clearpage
\subsection*{Supplementary Figures}

\begin{figure}[ht]
  \centering
      \includegraphics[width=\textwidth,height=\textheight,keepaspectratio]{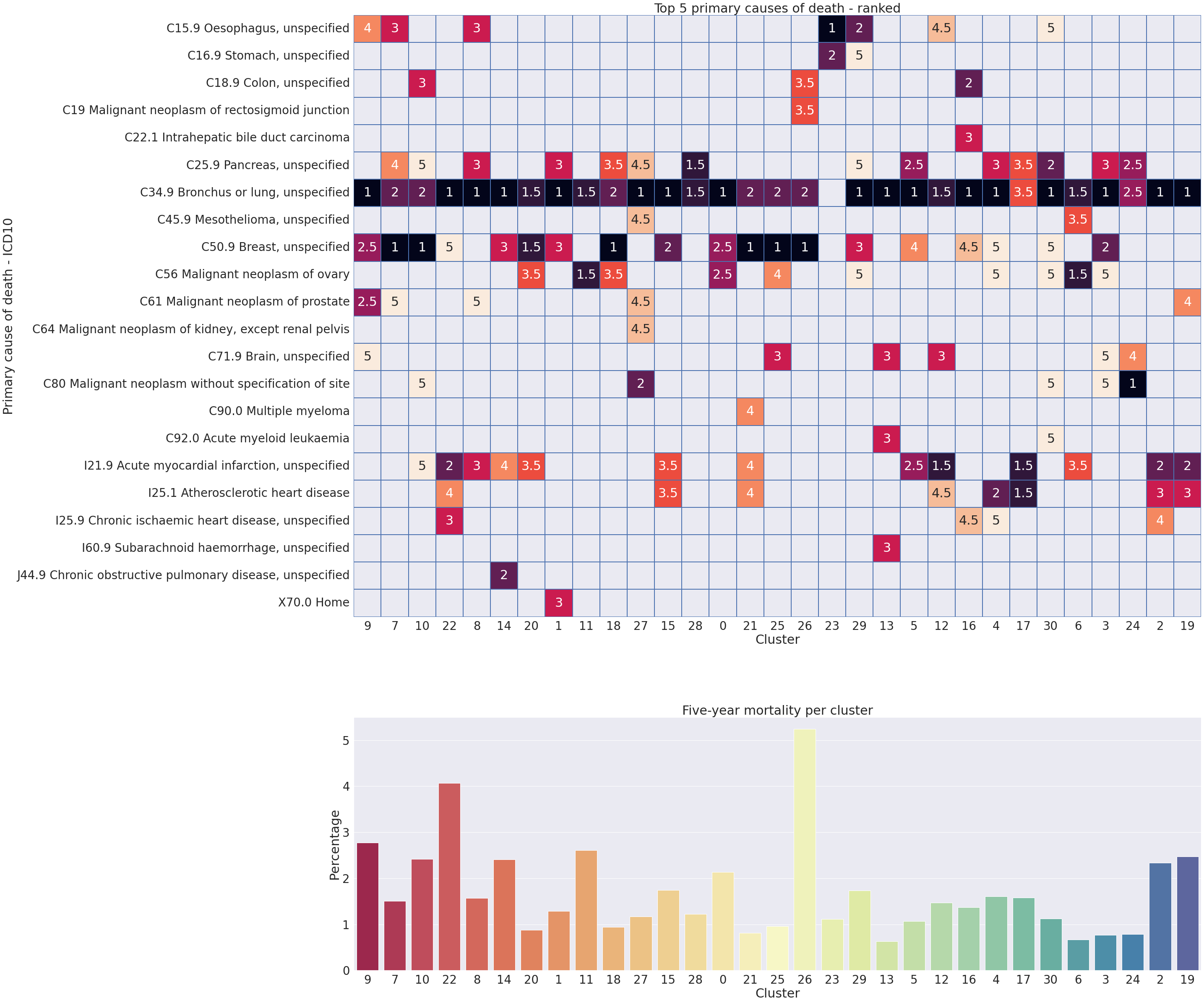}
      \caption{Primary cause of death and five-year mortality for the main clusters.\label{fig:c31-pcod-5ymort}}
\end{figure}

\begin{figure}[ht]
  \centering
      \includegraphics[width=\textwidth,height=\textheight,keepaspectratio]{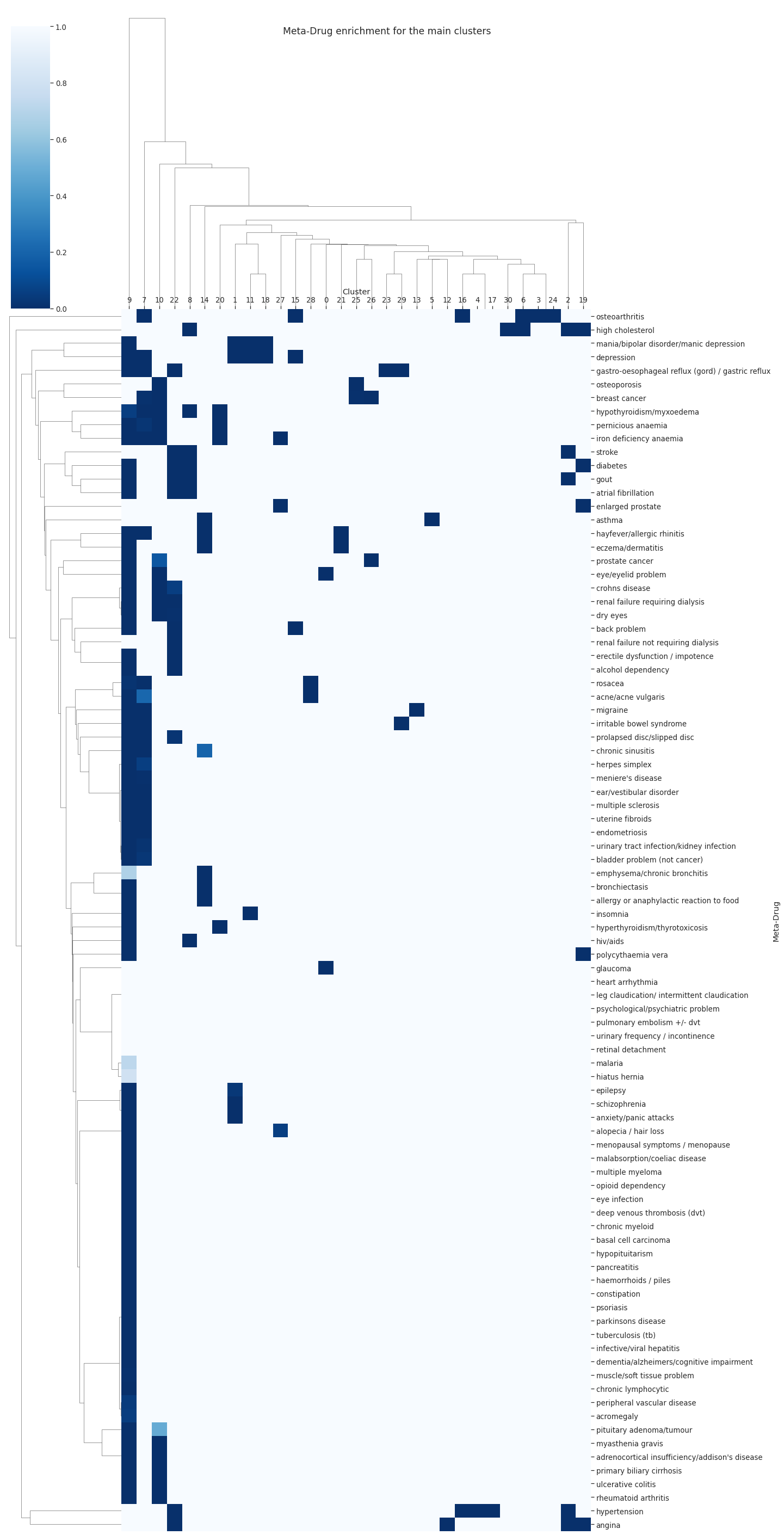}
      \caption{Meta-Drug enrichment for the main clusters.\label{fig:c31-metadrug-enrichment}}
\end{figure}

\begin{figure}[ht]
  \centering
      \includegraphics[width=\textwidth,height=\textheight,keepaspectratio]{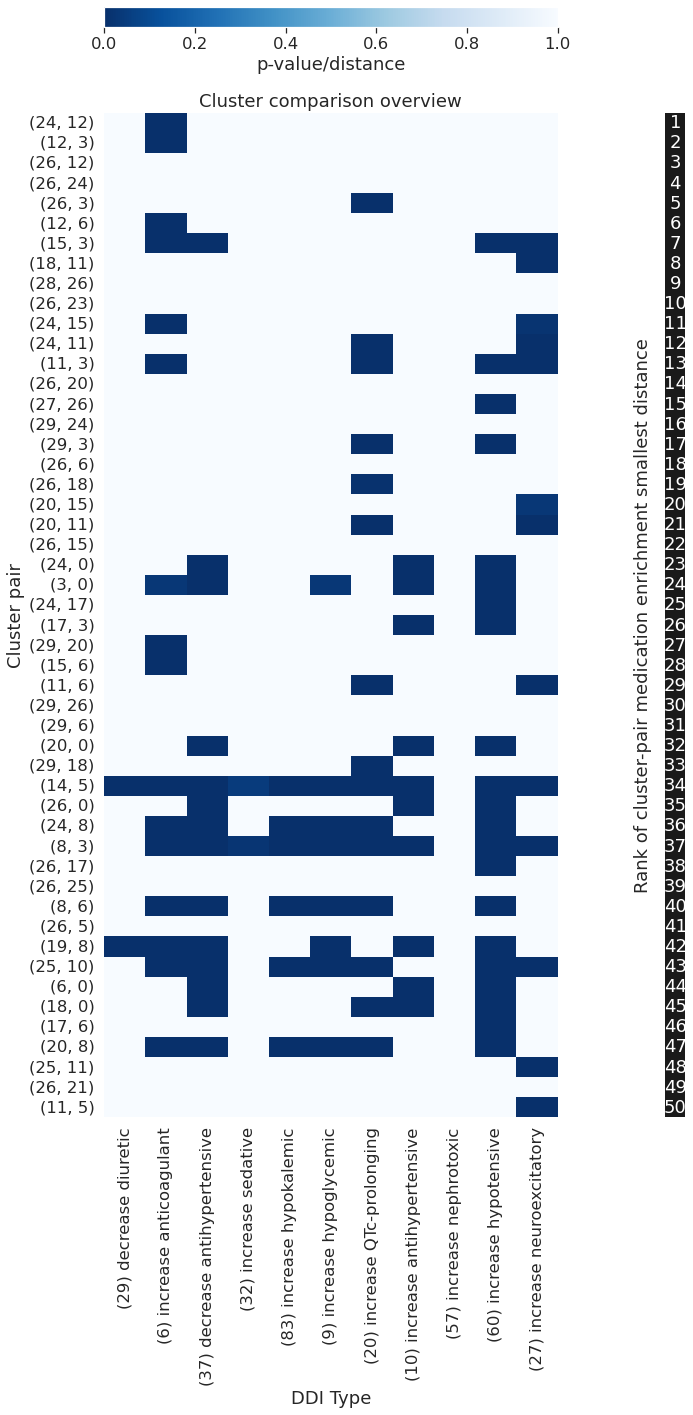}
      \caption{Cluster comparison for cluster pairs with significantly different five-year mortality - DDI types.\label{fig:c31-cluster-comparison-dditypes}}
\end{figure}

\end{document}